\documentclass[12pt,a4paper]{article}

\usepackage[english]{babel}
\usepackage{amsmath,amssymb,amsfonts,amsthm}
\usepackage[mathscr]{eucal}
\usepackage[all]{xy}
\usepackage{hyperref}
\usepackage{setspace}
\usepackage{upgreek}
\usepackage{hyperref}
\usepackage{cite}

\usepackage{times}
\usepackage{color}

\usepackage{indentfirst}

\allowdisplaybreaks

\begin{document}

\title{Extended Chern-Simons model for a vector multiplet}

\author {D.S.~Kaparulin\footnote{dsc@phys.tsu.ru} $\,{}^{1,2}$,
S.L.~Lyakhovich\footnote{sll@phys.tsu.ru} $\,{}^1$, and
O.D.~Nosyrev\footnote{nosyrevod@phys.tsu.ru} $\,{}^1$}
\date{${}^1$
\textit{Physics Faculty, Tomsk State University, Tomsk 634050,
Russia }\\
${}^2$\emph{Lebedev Institute of Physics, Leninsky ave. 53, Moscow
119991, Russia}}

\maketitle

\begin{abstract}We consider a gauge theory of vector fields in $3d$
Minkowski space. At the free level, the dynamical variables are
subjected to the extended Chern-Simons (ECS) equations with higher
derivatives. If the color index takes $n$ values, the third-order
model admits a $2n$-parameter series of second-rank conserved
tensors, which includes the canonical energy-momentum. Even though
the canonical energy is unbounded, the other representatives in the
series can have bounded from below $00$-component. The theory admits
consistent self-interactions with the Yang-Mills gauge symmetry. The
Lagrangian couplings preserve the unbounded from below
energy-momentum tensor, and they do not lead to a stable non-linear
theory. The non-Lagrangian couplings are consistent with the
existence of conserved tensor with a bounded from below
$00$-component. These models are stable at the non-linear level. The
dynamics of interacting theory admits a constraint Hamiltonian form.
The Hamiltonian density is given by the $00$-component of the
conserved tensor. In the case of stable interactions, the Poisson
bracket and Hamiltonian do not follow from the canonical
Ostrogradski construction. The particular attention is paid to the
"triply massless" ECS theory, which demonstrates instability already
at the free level. It is shown that the introduction of extra scalar
field, serving as Higgs, can stabilize dynamics in the vicinity of
the local minimum of energy. The equations of motion of stable model
are non-Lagrangian, but they admit the Hamiltonian form of dynamics
with a bounded from below Hamiltonian.\end{abstract}

\section{Introduction}

The higher derivative theories are notorious for better convergency
properties at classical and quantum level, and a wider symmetry. In
most of the instances, these advantages come with a price of
dynamics instability, being the typical problem for the models in
this class. At the classical level the solutions to the equations of
motion demonstrate runaway behavior ("collapse"). At the quantum
level the ghost poles appear in the propagator, so the unitarity of
dynamics is an issue. These peculiarities follow from a single fact:
the canonical energy is unbounded in every non-singular higher
derivative theory. For a review of the problem, we cite recent
articles \cite{Smilga, Pavsic, Boulanger} and references therein.
The canonical energy of singular models can be bounded. The examples
include $f(R)$-theories of gravity \cite{fR1, fR2, Tomboulis,
Belenchia}. These models do not demonstrate instability. The
stability problem for constrained higher derivative theories is
discussed in \cite{Chen}. In the majority of interesting
(constrained or unconstrained) models, the instability has the
special form: the classical dynamics is regular (no "collapse"), but
the canonical energy is unbounded from below. The Pais-Uhlenbeck
oscillator \cite{PU}, Podolsky \cite{Pod} and Lee-Wick \cite{LW1,
LW2} electrodynamics, ECS theory \cite{Deser}, and conformal gravity
\cite{Weyl} are examples. For the current studies we mention
\cite{FJ, Manavella, Dai, Pope1} and references therein. In all
these models the quantum theory is expected to be well-defined
\cite{Smilga, Smilga1}, but the application of the canonical
quantization procedure based on the Ostrogradski
procedure\footnote{The canonical Hamiltonian formalism for the
non-singular Higher-derivative theories has been first proposed in
\cite{Ostr}. Its generalization for singular theories has been
developed in \cite{GLT}: see also \cite{BL}. For recent studies see
\cite{KOT,OE}.} leads to the model with an unbounded from below
spectrum of energy. That is why the stability and unitarity of
quantum dynamics is the most important issue.

Recently, it has been recognized that the higher derivative dynamics
can be stable at the classical and quantum levels even if the
canonical energy of the model is unbounded. The articles
\cite{BK,DS} explain the stability of Pais-Uhlenbeck theory form the
viewpoint of existence of the Hamiltonian form of dynamics with a
bounded (non-canonical) Hamiltonian. The quantization of the model
with an alternative Hamiltonian leads to quantum theory with a
well-defined vacuum state with the lowest energy. The papers
\cite{BenM, Most} use a special PT-symmetry for construction of
stable quantum mechanics in the higher derivative oscillator model.
The articles \cite{Smilga, Pavsic, Boulanger, PUPert} tell us that
the non-linear higher derivative models may have a well-defined
classical dynamics without "collapsing" trajectories with the
runaway behavior. In the last articles, the existence of stable
classical dynamics serves as a necessary prerequisite for
construction of a well-defined quantum theory. For the studies of
field theories, we refer to articles \cite{Pope, Salvio}. All the
mentioned in this paragraph models have one common feature: they
admit an alternative (different from the canonical energy) bounded
from below conserved quantities. The quantum stability is achieved
if the model admits Hamiltonian form of dynamics, with the bounded
conserved quantity being the Hamiltonian. This means that the stable
higher derivative theory is characterized by two ingredients: the
bounded conserved quantity, and the Poisson bracket on its phase
space that brings the equations of motion to the Hamiltonian form
with a bounded from below Hamiltonian.

In the article \cite{KLS14}, the stability is studied in the class
of models of derived type. At the free level, the wave operator of
the theory is given by a polynomial (characteristic polynomial) in
another formally self-adjoint operator of lower order. It has been
shown that each derived theory admits a series of conserved tensors,
which includes the canonical energy-momentum \cite{KKL}. The number
of entries in the series grows with the order of characteristic
polynomial. Even though the canonical energy is unbounded, the other
conserved tensors in the series can be bounded from below
\cite{KKL}. The bounded conserved quantity stabilizes the classical
dynamics of the theory. The quantum stability is explained by the
existence of the Lagrange anchor that relates a bounded conserved
quantity with the invariance of the model with respect to the time
translations.\footnote{The concept of the Lagrange anchor has been
proposed in \cite{KazLS} in the context of quantization of
non-Lagrangian field theories. Later, it has been recognized that
the Lagrange anchor connects symmetries and conserved quantities
\cite{KLS10}.} In the first-order formalism, the Lagrange anchor
defines the Poisson bracket \cite{KLS-m}. The Hamiltonian is given
by the conserved quantity in the series, being expressed in terms of
phase-space variables. The linear higher derivative theory is stable
if a bounded conserved quantity, being related to the
time-translation symmetry, is admitted by the model \cite{KLS14}.
The articles \cite{KLS16, AKL19} consider a problem of consistent
deformations of symmetries and conserved quantities that preserve
the stability of dynamics in the class of derived type models. In
all the cases interaction vertices do not come from the least action
principle with higher derivatives, but the equations of motion admit
the Hamiltonian form of dynamics. The main difficulty of the cited
above procedures is that they do not automatically preserve gauge
invariance. This restricts their applications in gauge systems.

The ECS model \cite{Deser} is the simplest gauge theory of derived
type. In the current studies, it is often serves as a prototype of
the class of gauge theories with higher derivatives, see eg.
\cite{Sararu, Bufalo, Petrov}. The stability of the ECS model has
been first studied in \cite{KKL}. It has been observed that the
theory of order $p$ admits an $p-1$-parameter series of second-rank
conserved tensors, which includes the canonical energy-momentum. The
canonical energy of the model is always unbounded from below. The
other tensors in the series may have a bounded from below
00-component. The stability conditions for the theory are determined
by the structure of the roots of characteristic polynomial
\cite{KKL}. The theory is stable if all the nonzero roots of
characteristic polynomial are real and different and zero root has
multiplicity one or two. The stability of the ECS theory is
confirmed in work \cite{Dai1}. The ECS model has been shown to be a
multi-Hamiltonian at the free level in \cite{AKL18}. The series of
Hamiltonians includes the canonical (Ostrogradski) Hamiltonian,
which is unbounded, and the other representatives, being bounded or
unbounded depending on the model parameters. If the Hamiltonian is
bounded, it ensures the stability of the model at both classical and
quantum levels. The model admits inclusion of stable non-Lagrangian
interactions with scalar, fermionic and gravitational fields that
preserve a selected representative in the series of conserved
quantities of free model \cite{AKL18, KKL18, AKL19}. However, the
gauge symmetry is abelian in the sector of vector field in all these
examples.

The concept of consistency of interaction between the gauge fields
has been developed in \cite{Henn1}. The consistent couplings between
gauge vectors has been studied in \cite{Henn2, Henn3}. It has been
shown that the Yang-Mills vertex is unique in the covariant setting.
The Lagrangian self-interactions of the gauge vector multiplet
subjected to the ECS equations at free level has been reconsidered
in article \cite{Dai1}. The most general consistent non-linear
theory is proven to have the Yang-Mills gauge symmetry, and the
Lagrangian is given by the covariantization of the free ECS
Lagrangian. The interacting model demonstrates the Ostrogradski
instability at the non-linear level in all the instances, even
though the free theory is stable. This result implies a no-go
theorem for stable Lagrangian interactions in the ECS model. The
conclusion is not surprising because the Lagrangian couplings
preserve the conservation law of canonical energy, being unbounded
already at the free level. The inclusion of non-Lagrangian
interactions can solve the issue of the dynamics stability at the
interacting level, because such couplings preserve bounded conserved
quantities. However, the problem of construction of stable
consistent non-Lagrangian (self-)couplings (stable or unstable) has
not been studied in the ECS model in the literature before.

In the current work, we present a class of stable self-interactions
in the theory of vector multiplet. The free fields are subjected to
the ECS equations of third order. The interaction is (in general)
non-Lagrangian. The non-linear equations of motion are consistent
with the Yang-Mills gauge symmetry. A selected second-rank conserved
tensor of the free model is preserved by the coupling. Depending on
the values of coupling constants, it can be bounded or unbounded.
The Lagrangian interaction vertex of the article \cite{Dai1} is
unstable. The non-Lagrangian coupling can be consistent with the
existence of bounded conserved tensor. The equations of motion admit
the Hamiltonian form of dynamics. On shell, the Hamiltonian density
is given by the $00$-component of the conserved tensor. For
Lagrangian interactions, the canonical formalism with the unbounded
Ostrogradski Hamiltonian is reproduced. The bounded Hamiltonians do
not follow from the Ostrograski procedure, so we have non-canonical
Hamiltonian formalism in this case. In all the instances, the
Poisson bracket is a non-degenerate tensor, so the model admits a
Hamiltonian action principle. The quantization of the first-order
action with the bounded Hamiltonian leads to a stable quantum theory
with a well-defined vacuum state.

In the case of resonance (multiple roots of charlatanistic
polynomial), the dynamics of the theory is unstable already at the
free level. The inclusion of self-interaction does not improve
stability properties of dynamics of this model because the conserved
quantities have one and the same structure in the free and
non-linear models. In the current article, a model with the
third-order zero resonance root is of interest. The free field is
subjected to the "triply massless" Chern-Simons (CS) equations. The
wave operator of free equations is given by a cube of the CS
operator. To stabilize the dynamics at the non-linear level, we
apply the "Higgs-like" mechanism described in \cite{KLN20}.
Introducing an extra scalar, we generate a coupling such that the
energy of the interacting theory gets a local minimum for a solution
with nonzero values of dynamical variables. The motions of small
fluctuations in the vicinity of this solution are stable because the
non-linear theory has no resonance. The dynamics of fluctuations
admits the Hamiltonian form, with the Hamiltonian being given by the
positive definite quadratic form of the fields. This means that the
dynamics of the theory with a resonance can be stabilized by the
inclusion of appropriate interaction.

The article is organized as the following. In Section 2, we consider
the third-order ECS model for a vector multiplet. The particular
attention is paid to the structure of symmetries and conserved
quantities of the theory, and the stability of dynamics. In Section
3, we propose stable self-interactions between the multiplet of
vector fields with the Yang-Mills gauge symmetry. The general
non-model in the class model preserve a selected conserved tensor of
the free theory, which can be bounded or unbounded from below
depending on the values of coupling constants. The stable
interactions are inevitably non-Lagrangian. In Section 4, we
construct the constrained Hamiltonian formalism for the non-linear
model. The density of Hamiltonian is given by the 00-component of
the conserved tensor, being expressed in terms of the phase-space
variables. In Section 5, we consider the issue of stability of
"triply massless" ECS theory. We propose the class of consistent
couplings with the scalar field stabilizing the dynamics in the
vicinity of equilibrium position. The equations of motion are
non-Lagrangian, but they admit the constrained Hamiltonian form with
a bounded Hamiltonian. The concluding section summarizes the results
of the article.

\section{Higher-derivative Chern-Simons model}

We consider the most general third-order gauge theory of the vector
multiplet $A{}^a=A{}^a{}_\mu(x)dx^\mu$, $\mu=0,1,2$,
$a=1,2,\ldots,n$ on three-dimensional Minkowski space. The action
functional of the model reads
\begin{equation}\label{S-free}
    S[A(x)]=\frac{1}{2}\int A^a{}_\mu(\alpha_1
    F^{a\mu}+\alpha_2G^{a\mu}+\alpha_3K^{a\mu})d^3x\,.
\end{equation}
The real numbers $\alpha_1$, $\alpha_2$, $\alpha_3$ are model
parameters. Without loss of generality, we assume that the
coefficient at the higher derivative term is nonzero,
$\alpha_3\neq0$. The vectors $F{}^a{}_\mu$, $G{}^a{}_\mu$,
$K{}^a{}_\mu$ denote the (generalized) field strengthes of the
potential $A{}^a{}_\mu$,
\begin{equation}\label{FKG}
 F{}^a{}_{\mu}=\varepsilon_{\mu\nu\rho}\partial^\nu
    A{}^{a\rho}\,,\qquad G{}^a{}_{\mu}=\varepsilon_{\mu\nu\rho}\partial^\nu
    F{}^{a\rho}\,, \qquad K{}^a{}_{\mu}=\varepsilon_{\mu\nu\rho}\partial^\nu
    G{}^{a\rho}\,,
\end{equation}
with the Levi-Civita symbol $\epsilon_{\mu\nu\rho},
\epsilon_{012}=1$ being antisymmetric with respect to indices.  All
the tensor indices are raised and lowered by the Minkowski metric.
We use the mostly positive convention for the signature of metric
throughout the paper. The summation is implied over repeated at one
level isotopic indices $a=1,\ldots,n$ unless otherwise stated. For
$\alpha_3=\alpha_1=1$, $\alpha_2=0$, $n=1$, the action functional
(\ref{S-free}) has been first proposed in the article \cite{Deser}.
We call the model (\ref{S-free}) the ECS theory for a vector
multiplet.

The least action principle for the functional (\ref{S-free}) brings
us to the following Lagrange equations for the field $A_\mu$:
\begin{equation}\label{EL-free}
    \frac{\delta S}{\delta A^{a\mu}}\equiv \alpha_1
    F{}^{a}{}_\mu+\alpha_2G{}^a{}_\mu+\alpha_3K{}^a{}_\mu=0\,,\qquad
    a=1,\ldots,n\,.
\end{equation}
These equations have the derived form \cite{KKL} because the wave
operator is a polynomial in the CS operator $\ast d$. The symbol
$\ast$ denotes the Hodge star operator, and $d$ is the de Rham
differential. The structure of the Poincare group representation,
being described by the equation (\ref{EL-free}), is determined by
the roots of the characteristic polynomial
\begin{equation}\label{cp}
    M(\alpha;z)=\alpha_1z+\alpha_2z^2+\alpha_3z^3\,.
\end{equation}
The polynomial $M(\alpha;z)$ follows from (\ref{EL-free}) after the
formal replacement of the CS operator $\ast d$ by the complex-valued
variable $z$. The following cases are distinguished in \cite{KKL}.

(1a) $\alpha_1\neq0\,, \alpha{}_2{}^2-4\alpha_1\alpha_3>0\,$. The
characteristic polynomial has a zero root, and two different nonzero
real roots. Equation (\ref{EL-free}) describes two massive spin-1
self-dual fields.\footnote{A theory of massive spin-1 field
subjected to the self-duality condition has been proposed in
\cite{Townsend}. For the theory of representations of $3d$ Poincare
group, we cite \cite{Binegar, Grigore1, Grigore2}.} A zero root
corresponds to the CS mode, being pure gauge.

(1b) $\alpha_1\neq0\,,\alpha_2^2-4\alpha_1\alpha_3<0$. The
characteristic polynomial has a zero root and two complex roots. The
case is similar to (1a), but the masses of vector modes are complex.
The Poincare group representation is non-unitary.

(2a) $\alpha_2\neq0,\alpha_1=0$. The characteristic polynomial has a
multiplicity two zero root and simple nonzero root. The set of
subrepresentations includes a massless spin-1 field and a massive
spin-1 mode subjected to the self-duality condition.

(2b) $\alpha_2\neq0, \alpha_2^2-4\alpha_1\alpha_3=0$. The
characteristic polynomial has a multiplicity two nonzero root, and a
simple zero root. The subrepresentations describe "doubly massive"
massive mode, and a gauge CS mode. The Poincare group representation
is non-unitary.

(3)  $\alpha_1=\alpha_2=0$. The characteristic polynomial has a
multiplicity three zero root. Equation (\ref{EL-free}) describes the
"triply" massless extended CS theory. The Poincare group
representation is non-unitary.

As we see, the ECS model describes the unitary representation of the
Poincare group if all the nonzero roots of characteristic equation
are different, and the zero root has multiplicity one or two. The
dynamical degrees of freedom include the massive spin 1 vector
subjected to the self-duality condition and/or spin 1 massless
field, which meets the $3d$ Maxwell equations. In all the instances,
the model has two local physical degrees of freedom (four physical
polarizations).

The action functional (\ref{S-free}) is preserved by the
$2n$-parameter series of infinitesimal transformations
\begin{equation}\label{S-sym}
    \delta_{\xi;\beta} A{}^a{}_\mu=-\varepsilon_{\mu\nu\rho}\xi{}^{\rho}(\beta{}^a{}_1
    F{}^a{}^\rho+\beta{}^a{}_2G{}^a{}^\rho)\,,\qquad a=1,\ldots,n,
\end{equation}
(no summation in $a$). The transformation parameters are the
constant vector $\xi{}^{\mu}$ and real numbers $\beta{}^a{}_k$,
$a=0,\ldots,n$, $k=1,2$. The series (\ref{S-sym}) includes the
space-time translations with the independent parameters for the
individual vector $A{}^a{}_\mu$ in the multiplet, and higher order
transformations, whose value is determined by $\beta{}^a{}_2$. The
Noether theorem associates symmetries (\ref{S-sym}) with the
$2n$-parameter series of second-rank conserved tensors, which has
the form
\begin{equation}\label{Theta-ab}\begin{array}{c}\displaystyle
    \Theta^{\mu\nu}(\beta;\alpha)=\sum_{a=1}^n(\beta{}^a{}_1\Theta{}^{a\mu\nu}{}_1(\alpha)+
    \beta{}^a{}_2\Theta{}^{a\mu\nu}{}_2(\alpha))\,,
\end{array}\end{equation}
where
\begin{equation}\label{Theta-a-can}\begin{array}{c}\displaystyle
    \Theta{}^{a\mu\nu}{}_1(\alpha)=\alpha_3(G{}^{a\mu} F{}^{a\nu}+
    G{}^{a\nu} F{}^{a\mu}-g^{\mu\nu}G{}^a{}_\rho F{}^{a\rho})+\alpha_2(F{}^{a\mu} F{}^{a\nu}-\frac{1}{2}g^{\mu\nu}F{}^a{}_\rho
    F{}^{a\rho})\,;
\end{array}\end{equation}
\begin{equation}\label{Theta-a-add}\begin{array}{c}\displaystyle
    \Theta{}^{a\mu\nu}{}_2(\alpha)=\alpha_3(G{}^{a\mu}
    G^{a\nu}-\frac{1}{2}g^{\mu\nu}
    G{}^a{}_\rho G{}^{a\rho})-\alpha_1(F{}^{a\mu} F{}^{a\nu}-\frac{1}{2}g^{\mu\nu}F{}^a{}_\rho
    F{}^{a\rho})\,
\end{array}\end{equation}
(no summation in $a$). The quantities $\Theta{}^{a\mu\nu}{}_1$,
$a=1,\ldots,n$ (\ref{Theta-a-can}) represent the canonical
energy-momentum tensors of individual fields in the vector multiplet
$A{}^a{}_\mu$. The tensors $\Theta{}^{a\mu\nu}{}_2$, $a=1,\ldots,n$
(\ref{Theta-a-add}) are other conserved quantities. The total number
of independent conserved tensors in the free theory is $2n$ because
each field in the multiplet admits two different symmetries.

The 00-component of the conserved tensor (\ref{Theta-ab}) reads
\begin{equation}\label{Theta-ab-00}\begin{array}{c}\displaystyle
    \Theta^{00}(\beta;\alpha)=\frac{1}{2}
    \sum_{a=1}^n\Big\{\beta_2\alpha_3(G{}^{a0}G{}^{a0}
    +G{}^{ai}G{}^{ai})+2\beta_1\alpha_3
    (G{}^{a0}F{}^{a0}+G{}^{ai}F{}^{ai})+\\[5mm]\displaystyle+(\beta_1\alpha_2-\beta_2\alpha_1)
    (F{}^{a0}F{}^{a0}+F{}^{ai}F{}^{ai})\Big\}\,.
\end{array}\end{equation}
The summation over repeated at one level index $i=1,2$ is implied.
The quadratic form (\ref{Theta-ab-00}) can be reduced to the
principal axes as follows
\begin{equation}\label{Theta-ab-00-pa}\begin{array}{c}\displaystyle
    \Theta^{00}(\beta;\alpha)=\sum_{a=1}^{n}\Big\{\frac{\alpha_3(X{}^{a0}X{}^{a0}+X{}^{ai}X{}^{ai})}{2\beta{}^a{}_2}+
    \frac{ C(\beta^a;\alpha)(F{}^{a0}F{}^{a0}+F{}^{ai}F{}^{ai})}{2\beta{}^a{}_2}\Big\}\,.
\end{array}\end{equation}
Here, $X{}^a_\mu=\beta{}^a{}_1F{}^a{}_\mu+\beta{}^a{}_2G{}_\mu$, and
the notation is used
\begin{equation}\label{}
 C(\beta{}^a;\alpha)=-(\beta{}^a{}_2){}^2\alpha{}_1+\beta{}^a{}_2\beta{}^a{}_1\alpha_2-(\beta{}^a{}_1){}^2\alpha_3\,.
\end{equation}
In Section 4, we see that $X{}^{a0}X{}^{a0}+X{}^{ai}X{}^{ai}$ and
$F{}^{a0}F{}^{a0}+F{}^{ai}F{}^{ai}$ depend on different initial
data.\footnote{See the detailed discussion in \cite{AKL18}.} In this
case, the $00$-component (\ref{Theta-ab-00}) is bounded from below
if the coefficients at squares are positive,
\begin{equation}\label{Theta-pos}
    \phantom{\frac12}\beta{}^a{}_2\alpha{}_3>0,\qquad C(\beta{}^a;\alpha)>0\,,\qquad
    a=1,\ldots,n\,.\phantom{\frac12}
\end{equation}
We ignore the case of semi-positive quadratic form because the
degenerate conserved quantities do not ensure the stability of
dynamics. Relations (\ref{Theta-pos}) are consistent if and only if
the model parameters $\alpha_1,\alpha_2,\alpha_3$ meet conditions
(1a), (2a) of classification on pages 4-5. In the cases (1b), (2b),
(3) of this classification, conditions (\ref{Theta-pos}) are
inconsistent. This means that the free model (\ref{S-free}) is
stable if the wave equation (\ref{EL-free}) describes unitary
representation of the Poincare group. This result is quite natural
because the theories, which correspond to the unitary
representations of the Poincare group, should have a stable
dynamics.

\section{Stable interactions}

In this section, we present an example of stable self-interactions
in the model (\ref{S-free}). The interactions are non-Lagrangian.
The dynamics of the non-linear theory is determined by the equations
of motion. The interactions are associated with the deformations of
free equations of motion that preserve the gauge symmetries and
gauge identities of free model. The interaction is consistent if the
number of physical degrees of freedom is preserved by coupling. For
details of the concept of consistency of interaction in the class of
not necessary Lagrangian theories, we refer article \cite{KLS-inv}.

We start construction of non-linear theory by assuming that the
dynamical fields take values in the Lie algebra of a semisimple Lie
group with the generators $t{}^a,a=1,\ldots,n$,
\begin{equation}\label{}
    \mathcal{A}{}_\mu=A{}^a{}_\mu(x) t^a dx^\mu\,,\qquad
    [t{}^a,t{}^b]=f^{abc}t^c\,,\qquad
    \text{tr}(t{}^at{}^b)=\delta^{ab}\,.
\end{equation}
The covariant analogs of the (generalized) field strength vectors
(\ref{FKG}) are defined as follows:
\begin{equation}\label{FG}
    \mathcal{F}{}_\mu=\epsilon_{\mu\nu\rho}(\partial^\nu\mathcal{A}{}^\rho+
    \frac{1}{2}[\mathcal{A}{}^{\nu},\mathcal{A}{}^\rho])\,,\qquad
    \mathcal{G}_\mu=\epsilon_{\mu\nu\rho}D^\nu\mathcal{F}{}^\rho\,,\qquad
    \mathcal{K}_\mu=\epsilon_{\mu\nu\rho}D^\nu\mathcal{G}{}^\rho.
\end{equation}
The vectors $\mathcal{F}{}_\mu$, $\mathcal{G}{}_\mu$,
$\mathcal{K}{}_\mu$ lie in the Lie algebra of a semisimple Lie
group, and $D$ stands for the covariant derivative
\begin{equation}\label{}
    D_\mu=\partial_\mu+[\mathcal{A}_\mu,\,\cdot\,]\,.
\end{equation}
The vector $\mathcal{F}{}_\mu$ represents a dual of the standard
Yang-Mills strength tensor. The generalized field strengthes
$\mathcal{G}{}_\mu$, $\mathcal{K}{}_\mu$ are other covariant
quantities that involve second and third derivatives of
$\mathcal{A}{}_\mu$.

We consider the interactions that are polynomial in the gauge
invariants (\ref{FG}) and does not involve the highest derivative in
the non-linear part. In this setting, the most general
self-consistent non-linear theory is determined by the following
equations of motion:
\begin{equation}\label{EL-int}
    \mathbb{T}{}^\mu=\alpha{}_3\mathcal{K}{}^{\mu}+\alpha{}_2\mathcal{G}{}^{\mu}+
    \alpha{}_1\mathcal{F}{}^{\mu}-\frac{\alpha{}_3{}^2}{2 C(\beta;\alpha)}\epsilon^{\mu\nu\rho}
    [\beta{}_1\mathcal{F}{}_\nu+\beta{}_2\mathcal{G}{}_\nu,
    \beta{}_1\mathcal{F}{}_\rho+\beta{}_2\mathcal{G}{}_\rho]=0\,.
\end{equation}
We prove the uniqueness of this coupling in Appendix A. The model
parameters are the real numbers $\alpha_k,k=1,2,3$, $\beta_l$,
$l=1,2$, and $f^{abc},a,b,c=1,\ldots,n$. The constants
$\alpha{}_1,\alpha{}_2$ determine the free limit of the equations
(\ref{EL-int}). The numbers $\beta{}_1$, $\beta{}_2$ distinguish
admissible couplings with one and the same gauge group. Throughout
this section and below, we assume that $ C(\beta;\alpha)\neq0$. The
equations of motion (\ref{EL-int}) come from the least action
principle if $\beta_1=1,\beta_2=0$. In this case, the action
functional reads
\begin{equation}\label{S-int}
    S[\mathcal{A}(x)]=\frac{1}{2}\text{tr} \int \mathcal{F}{}_\mu(\alpha_1\mathcal{A}{}_\mu+
    \alpha_2\mathcal{F}{}^\mu+\alpha_3\mathcal{G}{}^\mu)d^3x\,.
\end{equation}
This action functional was first derived in
\cite{Dai1}.\footnote{The higher-derivative Yang-Mills theory having
a similar structure of the Lagrangian has been known long before
\cite{SlavFad}.} The same paper tells us that the (\ref{S-int}) is
the most general form of consistent self-interactions in the gauge
theory of vector fields. This means that the most general consistent
Lagrangian coupling (\ref{S-int}) is included into the model
(\ref{EL-int}). If the parameter $\beta_2$ is nonzero, equations
(\ref{EL-int}) do not follow from the least action principle for any
functional with higher derivatives. The variational principle with
auxiliary fields still exists, even if the higher derivative model
is non-Lagrangian. In the last case, the theory (\ref{EL-int})
admits consequent quantization, and establishing relationship
between symmetries and conserved quantities.

The concept of interaction consistency for not necessarily
Lagrangian theories has been developed in \cite{KLS-inv}. This paper
tells us that the non-Lagrangian interaction is consistent if the
non-linear theory admits the same number of (i) gauge symmetries,
(ii) gauge identities, and (iii) physical degrees of freedom as a
free model. All these fact are easily verified. At first, the
equations of motion (\ref{EL-int}) are preserved by the Yang-Mills
gauge symmetry,
\begin{equation}\label{gt-int}
    \phantom{\frac12}\delta_\zeta \mathcal{A}_\mu=D_\mu\zeta\,,\qquad
    \delta_\zeta
    \mathbb{T}_\mu=[\,\zeta\,,\mathbb{T}_\mu]\,,\phantom{\frac12}
\end{equation}
with $\zeta=(\zeta{}^a(x),a=1,\ldots,n)$ being the gauge
transformation parameter. The free model (\ref{S-int}) is preserved
by the standard gradient gauge symmetry, $\delta_\zeta
A{}^a{}_\mu=\partial_\mu\zeta^a$. As it is required, the gauge
symmetry (\ref{gt-int}) is given by the deformation of the gradient
gauge symmetry of free model. The important difference is that the
gauge symmetry (\ref{gt-int}) is non-abelian. So, the inclusion of
interaction leans to the model with non-abelian gauge symmetry. We
have an obvious set of gauge identities between the equations
(\ref{EL-int}),
\begin{equation}\label{gid-int}
    \phantom{\frac12}\mathbb{D}{}_\mu \mathbb{T}{}^{\mu}=0\,,\qquad
    \mathbb{D}_\mu=D_\mu+\frac{\alpha{}_3}{C(\beta;\alpha)}
    [\beta{}_1\mathcal{F}{}_\mu+\beta{}_2\mathcal{G}{}_\mu,\cdot]\,.\phantom{\frac12}
\end{equation}
Again, the leading term of the gauge identity is given by the free
contribution. This agrees with the concept of interaction
consistency. At the final step of analysis, we verify that the
physical degrees of freedom number is preserved by the coupling.
Equation (8) of paper \cite{KLS-inv} expresses the physical degrees
of freedom number via the orders gauge symmetries, gauge identities,
and equations of motion in the involutive of dynamics. The systems
(\ref{EL-free}), (\ref{EL-int}) are involutive, and they have equal
orders of gauge symmetries, gauge identities and equations of
motion. So, they have to poses one and the same physical degrees of
freedom number. All the above implies that the non-Lagrangian
interaction (\ref{EL-int}) is consistent for the general values of
the parameters $\beta,\alpha$.

The theory (\ref{EL-int}) admits a symmetric conserved tensor of
second rank of the following form:
\begin{equation}\label{Theta-ab-int}\begin{array}{c}\displaystyle
    \boldsymbol{\Theta}{}^{\mu\nu}{}(\beta;\alpha)=\text{tr}\Big\{\beta_2\alpha_3(\mathcal{G}{}^{\mu}
    \mathcal{G}^{\nu}-\frac{1}{2}g^{\mu\nu} \mathcal{G}{}_\rho \mathcal{G}{}^{\rho})+
    \beta_1\alpha_3(\mathcal{G}{}^{\mu} \mathcal{F}{}^{\nu}+
    \mathcal{G}{}^{\nu} \mathcal{F}{}^{\mu}-\\[5mm]\displaystyle-g^{\mu\nu}\mathcal{G}{}_\rho
    \mathcal{F}{}^{\rho})
    +(\beta_1\alpha_2-\beta_2\alpha_1)(\mathcal{F}{}^{\mu} \mathcal{F}{}^{\nu}-\frac{1}{2}g^{\mu\nu}\mathcal{F}{}_\rho
    \mathcal{F}{}^{\rho})\Big\}\,,
\end{array}\end{equation}
with $\alpha_k,k=1,2,3$ and $\beta_l,l=1,2$ being the model
parameters. The divergence of the quantity
$\boldsymbol{\Theta}{}^{\mu\nu}{}(\beta;\alpha)$ reads
\begin{equation}\label{}
    \partial_{\nu}\boldsymbol{\Theta}{}^{\mu\nu}{}=-\text{tr}(\epsilon^{\mu\nu\rho}(\beta{}_1\mathcal{F}{}_\nu+\beta{}_2\mathcal{G}{}_\nu)\mathbb{T}_\rho)\,.
\end{equation}
Expression (\ref{Theta-ab-int}) is the covariantization of a
selected representative in the conserved tensor series
(\ref{Theta-ab}),
\begin{equation}\label{ba-b}
    \phantom{\frac12}\beta{}^a{}_1=\beta_1\,,\qquad \beta{}^a{}_2=\beta_2\,,\qquad
    a=1,\ldots,n\,.\phantom{\frac12}
\end{equation}
As is seen, the model (\ref{EL-int}) represents the class of
deformations of free ECS equations (\ref{EL-free}) that preserves a
selected representative in the series (\ref{Theta-ab}) conserved
quantities at the non-linear level. It is important to note that the
other representatives in the series (\ref{Theta-ab-int}) no longer
conserve in the non-linear theory (\ref{EL-int}). This happens
because the parameters $\beta_k$, $k=1,2$ in the conserved tensor
(\ref{Theta-ab-int}) are unambiguously fixed by the interaction.

As far as stability is concerned, the $00$-component of tensor
(\ref{Theta-ab-int}) is relevant. The latter reads
\begin{equation}\label{Theta-ab-00-int}\begin{array}{c}\displaystyle
    \boldsymbol{\Theta}{}^{00}{}(\beta;\alpha)=\text{tr}\Big\{\frac{1}{2}\beta_2\alpha_3(\mathcal{G}{}^{0}
    \mathcal{G}^{0}+\mathcal{G}{}^i \mathcal{G}{}^{i})+
    \beta_1\alpha_3(\mathcal{G}{}^{0} \mathcal{F}{}^{0}+
    \mathcal{G}{}^{i} \mathcal{F}{}^{i})+\\[5mm]\displaystyle
    +\frac{1}{2}(\beta_1\alpha_2-\beta_2\alpha_1)
    (\mathcal{F}{}^{0} \mathcal{F}{}^{0}+\mathcal{F}{}^i
    \mathcal{F}{}^{i})\Big\}\,.
\end{array}\end{equation}
The conserved tensor is a bilinear form in (generalized) strengthes
$\mathcal{F}{}_\mu$, $\mathcal{G}{}_\mu$. The bringing of the
quadratic form (\ref{Theta-ab-00-int}) to the principal axes reveals
that the model is stable if
\begin{equation}\label{H-pos}
    \beta_2\alpha_3>0\,,\qquad  C(\beta;\alpha)>0\,.
\end{equation}
These conditions can be consistent or inconsistent depending on the
values of the model parameters $\alpha_k$, $k=1,2,3$ and $\beta_l$,
$l=1,2$. The Lagrangian interaction vertex (\ref{S-int}) does not
meet stability requirements. This confirms the instability of the
variational coupling proposed in \cite{Dai1}. The stable
interactions in the class of models (\ref{EL-int}) are inevitably
non-Lagrangian. The similar form of stability conditions
(\ref{EL-free}), (\ref{EL-int}) at the free and interacting cases
implies that the linear and non-linear dynamics is stable or
unstable simultaneously. In the class of stable at the linear level
theories, formula (\ref{EL-int}) determines a class of non-linear
models that preserve a selected bounded conserved quantity in the
series (\ref{Theta-ab}).

Now, we can return to the special case $ C(\beta;\alpha)=0$, which
is excluded in (\ref{EL-int}). The conserved quantity
(\ref{Theta-ab}), (\ref{ba-b}) is a degenerate quadratic form of the
initial data. The 00-component of the free conserved tensor can be
bounded from below, but its degeneracy implies existence of zero
vector(s). The motion of the system in the degenerate direction can
be infinite, and the corresponding conserved quantity appears to be
irrelevant to stability already at the free level. The formula
(\ref{EL-int}) prevents construction of interacting theories that
preserve the conserved tensors with a semi-definite 00-component.

\section{Hamiltonian formalism}

In this section, we construct constrained Hamiltonian formalism for
the higher-derivative theory (\ref{EL-int}).

Let us first explain what do we understand by the constrained
Hamiltonian formalism for the system of not necessarily Lagrangian
field equations with higher derivatives. A general fact is that the
higher-derivative system can be reduced to the first order by
introduction of extra fields absorbing the time derivatives of
original dynamical variables. The set of original dynamical
variables and extra fields is denoted by
$\{\varphi^I(\boldsymbol{\mathit{x}},t),\lambda^A(\boldsymbol{\mathit{x}},t)\}$,
where $t=x^0$ is the time, and $\boldsymbol{\mathit{x}}=(x^1,x^2)$
stands for space coordinates. The multi-indices $A,I$ label the
phase-space variables. The first-order equations are said to be
Hamiltonian if there exists a Hamiltonian function
$\mathcal{H}(\varphi^I,\boldsymbol{\nabla}\varphi^I,\boldsymbol{\nabla}^2\varphi^I,...)$
($\boldsymbol{\nabla}$ stands for derivatives by the space
$\boldsymbol{\mathit{x}}$) and a Poisson bracket
$\{\varphi^I(\boldsymbol{\mathit{x}}),\varphi^J(\boldsymbol{\mathit{y}})\}$
such that the equations constitute a constrained Hamiltonian system
(the dot denotes the derivative by time), i.e.
\begin{equation}\label{Ham-eq}\displaystyle\begin{array}{c}
    \dot{\varphi}^I(\boldsymbol{\mathit{x}})=\{\varphi^I(\boldsymbol{\mathit{x}}),\int
    \mathcal{H}d\boldsymbol{y}\,\},\quad
    \theta_A(\varphi^I(\boldsymbol{\mathit{x}}),\boldsymbol{\nabla}\varphi^I(\boldsymbol{\mathit{x}}),\ldots)=0\,;\\[5mm]
    \displaystyle
    \mathcal{H}=\mathcal{H}{}_0(\varphi^J(\boldsymbol{\mathit{x}}),\boldsymbol{\nabla}\varphi^J(\boldsymbol{\mathit{x}}),
    \ldots)+\lambda^A(\boldsymbol{x})\theta_A(\varphi^I(\boldsymbol{\mathit{x}}),\boldsymbol{\nabla}\varphi^I(\boldsymbol{\mathit{x}}),\ldots)\,.
\end{array}\end{equation}
The system is multi-Hamiltonian if there exists a series of
Hamiltonian $\mathcal{H}_\beta$ and Poisson brackets
$\{\varphi^I(\boldsymbol{\mathit{x}}),\varphi^J
(\boldsymbol{\mathit{y}})\}_{\beta}$ parameterized by constrains
$\beta_1,\ldots,\beta_k$ such that determine one and the same
equations of motion for the dynamical fields. The existence of the
constraint Hamiltonian formulation is not guaranteed for a system of
general non-Lagrangian equations of motion. In particular, the
Hamiltonian formulation for the ECS theory may not exist, at least
for certain combinations of the model parameters.

We start the construction of the Hamiltonian formalism from the
reduction of order of equations (\ref{EL-int}). The space components
of the field strength $\mathcal{F}_i,i=1,2,$ and generalized field
strength of second order $\mathcal{G}_i,i=1,2,$ are chosen as extra
fields. By construction, they absorb the first and second time
derivatives of space components of the vector field
$\mathcal{A}_i,i=1,2,$
\begin{equation}\label{FiGi}\displaystyle\begin{array}{c}
    \mathcal{F}{}_i=\epsilon_{ij}(\dot{\mathcal{A}}{}_j-\partial_i\mathcal{A}_0+[\mathcal{A}_0,\mathcal{A}_j])\,,\\[5mm]\displaystyle
    \mathcal{G}{}_i=-\ddot{\mathcal{A}}{}_i+\partial_i\dot{A}_0+
    \partial_j(\partial_j\mathcal{A}_i-\partial_i\mathcal{A}_j)+\epsilon_{ij}
    [\epsilon_{jk}\dot{\mathcal{A}}{}_0+\delta{}_{jk}(\epsilon_{rs}\partial{}_r\mathcal{A}{}_s+\\[5mm]\displaystyle
    +1/2[\mathcal{A}{}_r,\mathcal{A}{}_s],\mathcal{A}{}_i]+\epsilon_{ij}
    [\mathcal{A}{}_0,\epsilon{}_{jk}\dot{\mathcal{A}}{}_k+\delta_{jk}\mathcal{F}{}_k]\,.
\end{array}\end{equation}
Here, $\epsilon_{ij},\epsilon_{12}=1$ is the $2d$ Levi-Civita
symbol. The Latin indices $i,j$ run over the values $1,2$. The
summation over repeated at one and the same level Latin indices is
implied. As is seen form equations, (\ref{FiGi}) the fields
$\mathcal{F}{}_i$ absorb the first time derivatives of space
components of original vector field $\mathcal{A}{}_i$. The vector
$\mathcal{G}{}_i$ involves second time derivatives of
$\mathcal{A}{}_i$. The time components $\mathcal{F}_0,\mathcal{G}_0$
of vectors (\ref{FiGi}) are functions of $\mathcal{A}{}_0$,
$\mathcal{A}{}_i$, $\mathcal{F}{}_i$, $\mathcal{G}{}_i$, with no new
combinations of time derivatives being involved,
\begin{equation}\label{F0G0}\displaystyle\begin{array}{c}
    \phantom{\frac12}\mathcal{F}{}_0\equiv\epsilon_{ij}(\partial{}_i\mathcal{A}{}_j+[\mathcal{A}_i,\mathcal{A}_j])\,,\qquad
    \mathcal{G}{}_0\equiv\epsilon_{ij}(\partial_i\mathcal{F}_j+[\mathcal{A}_i,\mathcal{G}_j])\,. \phantom{\frac12}
\end{array}\end{equation}
In remaining part of the article, we associate the quantities
$\mathcal{F}_0$, $\mathcal{G}_0$ with their expressions in terms of
the phase-space variables $\mathcal{A}{}_0$, $\mathcal{A}{}_i$,
$\mathcal{F}{}_i$.

Substituting the extra variables (\ref{FiGi}) into the system
(\ref{EL-int}), we obtain the first-order equations for the fields
$\mathcal{A}_i$, $\mathcal{F}_i$, $\mathcal{G}_i$:
\begin{equation}\label{dotA}
    \phantom{\frac12}\dot{\mathcal{A}}{}_i=
    \partial_i\mathcal{A}_0-[\mathcal{A}{}_0,\mathcal{A}{}_i]-\epsilon_{ij}\mathcal{F}_j\,;\phantom{\frac12}
\end{equation}
\begin{equation}\label{dotF}
    \phantom{\frac12}\dot{\mathcal{F}}{}_i=\partial_i\mathcal{F}_0+
    [\mathcal{A}{}_i,\mathcal{F}{}_0]-[\mathcal{A}{}_0,\mathcal{F}{}_i]-
    \epsilon_{ij}\mathcal{G}_j\,;\phantom{\frac12}
\end{equation}
\begin{equation}\label{dotG}\begin{array}{c}\displaystyle
    \phantom{\sum_1^2}\dot{\mathcal{G}}{}_i=\partial_i\mathcal{G}_0+
    \frac{\alpha_2}{\alpha_3}\epsilon_{ij}\mathcal{G}_j+\frac{\alpha_1}{\alpha_3}\epsilon_{ij}\mathcal{F}_j+[\mathcal{A}{}_i,\mathcal{G}{}_0]-
    [\mathcal{A}{}_0,\mathcal{G}{}_i]+\phantom{\sum_1^2}\\[5mm]\displaystyle
    +\frac{\alpha{}_3{}^2}{ C(\beta;\alpha)}[\beta{}_1\mathcal{F}{}_0+\beta{}_2\mathcal{G}{}_0,\beta{}_1\mathcal{F}{}_i+\beta{}_2\mathcal{G}{}_i]\,.
\end{array}\end{equation} The evolutionary equations (\ref{dotA}),
(\ref{dotF}), (\ref{dotG}) are supplemented by the constraint
\begin{equation}\label{T-const}\begin{array}{c}\displaystyle
    \Theta=\epsilon_{ij}(\alpha_1\partial_i\mathcal{A}{}_j+\alpha_2\partial_i\mathcal{F}{}_j+
    \alpha_3\partial_i\mathcal{G}{}_j+[\mathcal{A}{}_i,\frac{1}{2}\alpha_1\mathcal{A}{}_j+\alpha{}_2\mathcal{F}{}_j+\alpha{}_3\mathcal{G}{}_j]+
    \\[5mm]\displaystyle-\frac{\alpha_3}{2 C(\beta;\alpha)}
    [\beta_1\mathcal{F}{}_i+\beta_2\mathcal{G}{}_i,\beta_1\mathcal{F}{}_j+\beta_2\mathcal{G}{}_j])\approx0\,.
\end{array}\end{equation}
The systems (\ref{EL-int}) and (\ref{dotA}), (\ref{dotF}),
(\ref{dotG}), (\ref{T-const}) are equivalent. Evolutionary equations
(\ref{dotA}), (\ref{dotF}) express the auxiliary fields
$\mathcal{F}{}_i, \mathcal{G}{}_i$ in terms of time derivatives of
the original vector potential $\mathcal{A}_\mu$. Solving them with
respect to the unknown $\mathcal{F}_i$, $\mathcal{G}_i$, we obtain
relations (\ref{FiGi}). Equations (\ref{dotG}), (\ref{T-const})
represent the space and time components of the original higher
derivative system (\ref{EL-int}) where all the higher derivatives of
the vector potential are expressed in terms of extra variables
$\mathcal{F}{}_i$, $\mathcal{G}{}_i$. Equation (\ref{T-const})
serves as constraint, which does not involve time derivatives. Once
it is preserved the time evolution (see the gauge identity
(\ref{gid-int})), no secondary constraints are imposed on the
fields.

We associate the on-shell Hamiltonian $\mathcal{H}_0$ with the
00-component of the conserved tensor (\ref{Theta-ab-int}). In terms
of the phase-space variables $\mathcal{A}{}_i$, $\mathcal{F}{}_i$,
$\mathcal{G}{}_i$, it reads
\begin{equation}\label{H-on}\begin{array}{c}\displaystyle
    \mathcal{H}_0=\frac{1}{2}\text{tr}\Big\{\beta_2\alpha_3
    (\mathcal{G}{}^0\mathcal{G}{}^0+\mathcal{G}{}^i\mathcal{G}{}^i)+
    2\beta_1\alpha_3(\mathcal{G}{}^0\mathcal{F}{}^0+\mathcal{G}{}^i\mathcal{F}{}^i)+
    \\[5mm]\displaystyle
    +(\beta_1\alpha_2-\beta_2\alpha_1)(\mathcal{F}{}^0\mathcal{F}{}^0+\mathcal{F}{}^i\mathcal{F}{}^i)\Big\}\,.
\end{array}\end{equation}
Here, the functions $\mathcal{F}{}^0$, $\mathcal{G}{}^0$ are defined
in (\ref{F0G0}). The on-shell Hamiltonian $\mathcal{H}_0$ depends on
the free model parameters $\alpha_k,k=1,2,3$ and coupling constants
$\beta_l,l=1,2$. Off-shell, The Hamiltonian is a sum of on-shell
part (\ref{H-on}) and a linear combination of constraints. We chose
the following ansatz for the total Hamiltonian,
\begin{equation}\label{H-off}\begin{array}{c}\displaystyle
    \mathcal{H}=\mathcal{H}_0+\text{tr}\Big[\frac{ C(\beta;\alpha)}{\beta{}_2\alpha{}_2-\beta{}_1\alpha{}_3}
    \mathcal{A}_0-\frac{\beta_2\beta_1\alpha_3}
    {\beta_2\alpha_2-\beta_1\alpha_3}\mathcal{F}_0-
    \frac{\beta_2{}^2\alpha_3}{\beta_2\alpha_2-\beta_1\alpha_3}\mathcal{G}_0\Big]\Theta\,,
\end{array}\end{equation}
where $\beta,\alpha$ are constant parameters. The on-shell vanishing
terms, being included into the Hamiltonian, does not contribute to
the equations of motion of gauge-invariant quantities. The equations
of motion do alter for non-gauge-invariant variables. As we are
seeking for the Hamiltonian and Poisson bracket literally
reproducing the first-order form (\ref{dotA}), (\ref{dotF}),
(\ref{dotG}), (\ref{T-const}) of equations (\ref{EL-int}), the
on-shell vanishing contributions are kept under control in
(\ref{H-off}).

The Hamiltonian is well-defined if the parameters $\beta,\alpha$
subject to the following conditions:
\begin{equation}\label{Delta-neq}
     C(\beta;\alpha)\neq0\,,\qquad
    \beta_1\alpha_3-\beta_2\alpha_2\neq0.
\end{equation}
These relations have a clear origin. The first condition in this set
insures that the on-shell Hamiltonian is a non-degenerate quadratic
form of the phase-space variables $\mathcal{A}{}_i$,
$\mathcal{F}{}_i$, $\mathcal{G}{}_i$. This requirement is reasonable
because the degenerate Hamiltonian cannot generate the evolution of
all physical degrees of freedom. The second relation
(\ref{Delta-neq}) ensures that the numerical factor at the Lagrange
multiplier $\mathcal{A}{}_0$ is non-singular. This is necessary to
reproduce the correct gauge transformations for all the dynamical
variables. We also note that the obstructions to existence of
Hamiltonian remain valid in the free limit. This means that the
inclusion of interaction by the scheme of Section 3 does not
restrict the class models that admit the Hamiltonian formulation. In
the other words, every theory in the class (\ref{EL-int}) admitting
the Hamiltonian formulation with the Hamiltonian (\ref{H-off}) in
the free limit, is Hamiltonian at the interacting level.
Hereinafter, we assume that the relations (\ref{Delta-neq}) are
satisfied. The Hamiltonian is on-shell bounded from below if the
conditions (\ref{H-pos}) are satisfied. In this case, we expect to
construct the constrained Hamiltonian formulation with a bounded
Hamiltonian. The corresponding quantum theory has a good chance to
be stable.

Now, let us seek for the Poisson bracket between the fields
$\mathcal{A}{}_i$, $\mathcal{F}{}_i$, $\mathcal{G}{}_i$ that brings
equations (\ref{dotA}), (\ref{dotF}), (\ref{dotG}), (\ref{T-const})
to the Hamiltonian form (\ref{Ham-eq}) with the Hamiltonian
(\ref{H-off}). Comparing the right hand sides of relations
(\ref{dotA}), (\ref{dotF}), (\ref{dotG}) and equations
(\ref{Ham-eq}), we obtain the following system of algebraic
equations:
\begin{equation}\label{PB-dotA}
    \phantom{\frac12}\Big\{A{}_i,\int \mathcal H(\boldsymbol{y})
    d\boldsymbol{y}\Big\}\approx
    \partial_i\mathcal{A}_0-[\mathcal{A}{}_0,\mathcal{A}{}_i]-\epsilon_{ij}\mathcal{F}_j\,;\phantom{\frac12}
\end{equation}
\begin{equation}\label{PB-dotF}
    \phantom{\frac12}\Big\{F{}_i,\int \mathcal H(\boldsymbol{y})
    d\boldsymbol{y}\Big\}\approx\partial_i\mathcal{F}_0+[\mathcal{A}{}_i,\mathcal{F}{}_0]-[\mathcal{A}{}_0,\mathcal{F}{}_i]-
    \epsilon_{ij}\mathcal{G}_j
    \,;\phantom{\frac12}
\end{equation}
\begin{equation}\label{PB-dotG}\begin{array}{c}\displaystyle
    \phantom{\sum_1^2}\Big\{G{}_i,\int \mathcal H(\boldsymbol{y}) d\boldsymbol{y}\Big\}\approx\partial_i\mathcal{G}_0+
    \frac{\alpha_2}{\alpha_3}\epsilon_{ij}\mathcal{G}_j+\frac{\alpha_1}{\alpha_3}\epsilon_{ij}\mathcal{F}_j
    +[\mathcal{A}{}_i,\mathcal{G}{}_0]-[\mathcal{A}{}_0,\mathcal{G}{}_i]
    +\phantom{\sum_1^2}\\[5mm]\displaystyle
    +\frac{\alpha{}_3{}^2}{C(\beta;\alpha)}[\beta{}_1\mathcal{F}{}_0+\beta{}_2\mathcal{G}{}_0,\beta{}_1\mathcal{F}{}_i+\beta{}_2\mathcal{G}{}_i]\,.
\end{array}\end{equation}
The sign $\approx$ means equality modulo constraint (\ref{T-const}).
The Poisson bracket, being defined by these equations, depends on
five independent arguments: the free model parameters $\alpha_3$,
$\alpha_2$, $\alpha_1$, and coupling constants $\beta_2$, $\beta_1$.
It has the following form:
\begin{equation}\label{PB-GG}
    \phantom{\sum_a^b}\{\mathcal{G}{}^a{}_i(\boldsymbol{x}),
    \mathcal{G}{}^b{}_j(\boldsymbol{y})\}=\frac{\beta_1\alpha_2{}^2-\beta_1\alpha_1\alpha_3-\beta_2\alpha_2\alpha_1}
    {\alpha{}_3{}^2C(\beta;\alpha)}\epsilon_{ij}\delta^{ab}
    \delta^{(2)}(\boldsymbol{x}-\boldsymbol{y})\,;\phantom{\sum_a^b}
\end{equation}
\begin{equation}\label{PB-FG}
    \phantom{\sum_a^b}\{\mathcal{F}{}^a{}_i(\boldsymbol{x}),
    \mathcal{G}{}^b{}_j(\boldsymbol{y})\}=\frac{\beta_2\alpha_1-\beta_1\alpha_2}
    {\alpha_3C(\beta;\alpha)}\epsilon_{ij}\delta^{ab}
    \delta^{(2)}(\boldsymbol{x}-\boldsymbol{y})\,;\phantom{\sum_a^b}
\end{equation}
\begin{equation}\label{PB-FF}
    \phantom{\sum_a^b}\{\mathcal{F}{}^a{}_i(\boldsymbol{x}),
    \mathcal{F}{}^b{}_j(\boldsymbol{y})\}=\{\mathcal{A}{}^a{}_i(\boldsymbol{x}),
    \mathcal{G}{}^b{}_j(\boldsymbol{y})\}=\frac{\beta_1}{C(\beta;\alpha)}\epsilon_{ij}\delta^{ab}
    \delta^{(2)}(\boldsymbol{x}-\boldsymbol{y})\,;\phantom{\sum_a^b}
\end{equation}
\begin{equation}\label{PB-AF}
    \phantom{\sum_a^b}\{\mathcal{A}{}^a{}_i(\boldsymbol{x}),
    \mathcal{F}{}^b{}_j(\boldsymbol{y})\}=\frac{-\beta_2}{C(\beta;\alpha)}\epsilon_{ij}\delta^{ab}
    \delta^{(2)}(\boldsymbol{x}-\boldsymbol{y})\,;\phantom{\sum_a^b}
\end{equation}
\begin{equation}\label{PB-AA}
    \phantom{\sum_a^b}\{\mathcal{A}{}^a{}_i(\boldsymbol{x}),
    \mathcal{A}{}^b{}_j(\boldsymbol{y})\}=0\,.\phantom{\sum_a^b}
\end{equation}
Here,
$\delta^{(2)}(\boldsymbol{x}-\boldsymbol{y})=\delta(x^1-y^1)\delta(x^2-y^2)$
is the $2d$ $\delta$-function in the space coordinates. The Poisson
bracket (\ref{PB-GG}), (\ref{PB-FG}), (\ref{PB-FF}), (\ref{PB-AF}),
(\ref{PB-AA}) is a covariant generalization of the its free analog,
being derived in \cite{AKL18}. This result is not surprising. The
free limit of expressions (\ref{PB-GG}), (\ref{PB-FG}),
(\ref{PB-FF}), (\ref{PB-AF}), (\ref{PB-AA}) is determined by the
linear model, while the Poisson brackets with the polynomial
dependence on fields contradicts the structure of equations
(\ref{Ham-eq}). In the last case, the Poisson bracket with the
Hamiltonian involves higher powers of fields than the right hand
side of first-order equations (\ref{dotA}), (\ref{dotF}),
(\ref{dotG}).

The Poisson bracket (\ref{PB-GG}), (\ref{PB-FG}), (\ref{PB-FF}),
(\ref{PB-AF}), (\ref{PB-AA}) is a non-degenerate tensor, so it has
inverse, being a symplectic two-form. The latter defines a
Hamiltonian action functional
\begin{equation}\label{S-h}\begin{array}{c}\displaystyle
    S_h=\int
    \Big\{\text{tr}\Big(\frac{ C(\beta;\alpha)}{\beta_2\alpha_2-\beta_1\alpha_3}
    \varepsilon_{ij}(\alpha_1\mathcal{A}{}_i+2\alpha_2\mathcal{F}{}_i+2\alpha_3\mathcal{G}{}_i)\dot{\mathcal{A}}{}_j+
    \frac{\beta_1{}^2\alpha_3}{\beta_2\alpha_2-\beta_1\alpha_3}\times\\[5mm]\displaystyle
    \times\varepsilon_{ij}
    (\beta{}_1\mathcal{F}{}_i+\beta{}_2\mathcal{G}{}_i)(\beta{}_1\mathcal{F}{}_j+\beta{}_2\mathcal{G}{}_j)\Big)
    -\mathcal{H}\Big\}d\boldsymbol{x}dt\,,
\end{array}\end{equation} where $\mathcal{H}$ denotes the Hamiltonian
(\ref{H-off}). In the case $\beta_2=0\,,\beta_1=1$, equation
(\ref{S-h}) reproduces the Ostrogradski action for the variational
model (\ref{S-int}),
\begin{equation}\label{S-c}
    S_{c}=\int
    \Big\{\text{tr}(\epsilon_{ij}(\alpha_1\mathcal{A}{}_i+2\alpha_2\mathcal{F}{}_i+
    2\alpha_3\mathcal{G}{}_i)\dot{\mathcal{A}}{}_j-\epsilon_{ij}
    \mathcal{F}{}_i\dot{\mathcal{F}}{}_j)-\mathcal{H}{}_{c}\Big\}d\boldsymbol{x}dt\,,
\end{equation}
where $\mathcal{H}_c$ is the canonical Hamiltonian,
\begin{equation}\label{}
    \mathcal{H}{}_c=\frac{1}{2}\text{tr}\Big(
    \alpha_3(\mathcal{G}{}^0\mathcal{F}{}^0+\mathcal{G}{}^i\mathcal{F}{}^i)+
    \alpha_2(\mathcal{F}{}^0\mathcal{F}{}^0+\mathcal{F}{}^i\mathcal{F}{}^i)+
    \mathcal{A}_0\Theta\Big)\,.
\end{equation}
In the free case, the Hamiltonian formulation (\ref{S-h}) has been
first proposed in \cite{AKL18}. Later, it has been re-derived in
\cite{Dai1} by the direct application of the Ostrogradski procedure
to the action functional (\ref{S-free}). The comparison of
conditions (\ref{Theta-pos}), (\ref{Delta-neq}) implies that the
Hamiltonian (\ref{H-off}) is on-shell unbounded for all the
Lagrangian interactions. For non-Lagrangian interactions, the
Hamiltonian (\ref{H-off}) can be on-shell bounded or unbounded
depending on the value of model parameters. If the Hamiltonian is
on-shell bounded, the interacting theory is stable at the quantum
level. To our knowledge, the ECS theory (\ref{EL-int}) is a first
higher derivative model with a non-abelian gauge symmetry admitting
alternative Hamiltonian formulation with a bounded Hamiltonian. This
means that the concept of stabilization of dynamics by means of
alternative Hamiltonian formalism applies beyond the linear level.

In the free limit, the Hamiltonian (\ref{H-off}) and Poisson bracket
(\ref{PB-GG}), (\ref{PB-FG}), (\ref{PB-FF}), (\ref{PB-AF}),
(\ref{PB-AA}) depend on the parameters $\beta_1$ and $\beta_2$,
while the equations of motion (\ref{dotA}), (\ref{dotF}),
(\ref{dotG}), (\ref{T-const}) do not. This means that the free ECS
theory (\ref{S-free}) is a multi-Hamiltonian theory. The general
representative the series of Hamiltonian formulations does not
follow from the Ostrogradski procedure. The reason is that the
bounded and unbounded Hamiltonian cannot be connected by the change
of coordinates in the phase space. The number of entries in the free
Hamiltonian series equals $2n$, where $n$ is the number of color
indices. The on-shell Hamiltonian is given by the $00$-component
(\ref{Theta-ab-00}) of the conserved tensor (\ref{Theta-ab}). The
Poisson bracket between the phase-space variables $A{}^a{}_i$,
$F{}^a{}_i$ ,$G{}^a{}_i$ is determined by the formula (\ref{PB-GG}),
(\ref{PB-FG}), (\ref{PB-FF}), (\ref{PB-AF}), (\ref{PB-AA}), but the
parameters $\beta_1$, $\beta{}_2$ are replaced by $\beta{}^a{}_1$,
$\beta{}^a{}_2$ for each field in the multiplet. At the non-linear
level, a selected Poisson bracket (\ref{PB-GG}), (\ref{PB-FG}),
(\ref{PB-FF}), (\ref{PB-AF}), (\ref{PB-AA}) is preserved. If the
respective Hamiltonian is bounded, the Poisson bracket cannot follow
from the Ostrogradski construction. So, the Hamiltonian formalism
for the stable non-linear theories does not follow from the
canonical formulation.

In the conclusion of this section, we present the Poisson brackets
between the constraints (\ref{PB-GG}), (\ref{PB-FG}), (\ref{PB-FF}),
(\ref{PB-AF}), (\ref{PB-AA}),
\begin{equation}\label{}
    \{\Theta{}^a(\boldsymbol{x}),\Theta{}^b(\boldsymbol{y})\}=
    \frac{\beta_2\alpha_2-\beta_1\alpha_3}{ C(\beta;\alpha)}
    f{}^{abc}\Theta{}^{c}(\boldsymbol{x})\delta(\boldsymbol{x}-\boldsymbol{y}).
\end{equation}
The Hamiltonian is gauge-invariant,
\begin{equation}\label{}
    \{\Theta{}^a(\boldsymbol{x}),
    \mathcal{H}{}_0(\boldsymbol{y})\}=0\,,\qquad a=1,\ldots,n\,.
\end{equation}
As is seen, the first-order model (\ref{S-h}) is gauge theory of
special form in the whole range (\ref{Delta-neq}) of admissible
values of the parameters $\alpha$, $\beta$. Equation (\ref{S-h})
determines the least action principle for the model. The
quantization of this theory can be performed by means of the
well-known procedures \cite{HT}. In all the instances with bounded
from below Hamiltonian, we expect the stability of the model. This
means that equations (\ref{EL-int}) determine a higher derivative
theory, which can be stable at classical and quantum level. Because
of existence of first-order formalism, the model is as good as the
Lagrangian theories. In particular, it admits consequent
quantization, and correspondence between symmetries and conserved
quantities. The Hamiltonian serves as quantity that is associated to
the invariance of the model (\ref{S-h}) with respect to the time
translations. The Hamiltonian (\ref{H-off}) is on-shell bounded if
the conditions (\ref{H-pos}) are satisfied.

\section{Resonance case}

In case of resonance (positions (2b), (3) of classification on pages
4-5), the non-linear theory (\ref{EL-free}) admits the Hamiltonian
form of dynamics, but the Hamiltonian is unbounded in all the
instances (conditions (\ref{H-pos}) are inconsistent). So, the model
has to be considered as unstable. In this section, we demonstrate
that the dynamics of the theory with resonance can be stabilized by
means of inclusion of interaction with extra dynamical field. We
apply the "Higgs-like" mechanism, which has been first proposed in
the context of study of the "doubly massless" generalized Podolsky
electrodynamics in the paper \cite{KLN20}. Here, we use it in the
theory with non-abelian gauge symmetry for the first time. We mostly
consider the model (\ref{EL-int}) with the third-order resonance. We
chose the following values for the parameters of free theory
(\ref{S-free}): $\alpha_1=\alpha_2=0$, $\alpha_3=-1$. This choice
does not restrict generality because the constant $\alpha_3$
accounts the possibility of multiplication of equations of motion by
an overall factor. The wave operator of free model (\ref{EL-free})
appears to be cube of the CS operator, so we have a sort of "triply
massless" extended theory.

We start construction of interaction by extending the set of
dynamical variables by a real scalar field $\phi(x)$. The non-linear
theory of vector multiplet $\mathcal{A}{}_\mu(x)$ and scalar field
$\phi(x)$ is determined by the equations of motion, which have the
following form:

\begin{equation}\label{ELF-int}\begin{array}{c}\displaystyle
    \mathbb{T}_\mu=\epsilon_{\mu\nu\rho}\Big\{D{}^\nu[(\widetilde{\gamma}{}^2\phi^2-1)\mathcal{G}{}+
    \gamma^2\phi^2\mathcal{F}]{}^\rho+
    \frac{(\widetilde{\gamma}{}^2\phi^2-1)^2}{2\beta_1{}^2}[\beta_1\mathcal{F}^{\nu}+\beta_2\mathcal{G}^\nu,\beta_1\mathcal{F}^{\rho}+\beta_2\mathcal{G}^\rho]\Big\}=0\,,\\[5mm]
    \displaystyle\mathbb{T}=\Big\{\partial_\mu\partial^\mu-(m^2-\widetilde{\gamma}^2\frac{(\beta{}_1\mathcal{F}_{\rho}+\beta{}_2\mathcal{G}_\rho)(\beta{}_1\mathcal{F}^{\rho}+\beta{}_2\mathcal{G}^\rho)}{\beta_1{}^2}
    +\phi^2\Big\}\phi=0\,.
\end{array}\end{equation}
Here, the vectors $\mathcal{F}{}_\mu$, $\mathcal{G}{}_\mu$ are
defined in (\ref{FKG}), and the abbrevialtion
$\widetilde{\gamma}{}^2=\gamma{}^2\beta_2/\beta_1$ is used. The
constants $\beta_2$, $\beta_1$, $\gamma$, $m$ are model parameters,
being real numbers. Throughout the section we assume that
$m,\gamma>0$ and $\beta_1\neq0$. The option $\beta_1=0$ is not
admissible because the self-coupling (\ref{EL-int}) between the
vector fields becomes inconsistent in this case. The value $\beta_2$
can be arbitrary real number (positive, negative or zero). Only
positive values of $\beta_2$, $\beta_1$ lead to stable couplings.
This justifies notation $\widetilde{\gamma}{}^2$ for the quantity
$\gamma^2\beta_2/\beta_1$. Without loss of generality, we put a unit
coefficient at $\phi^3$-term. An overall factor at the
$\phi{}^3$-vertex can be absorbed by the scaling of scalar field,
$\phi\mapsto\lambda\phi$ with the appropriate $\lambda\neq0$.

Equations (\ref{ELF-int}) have a clear meaning. The first line of
the system (\ref{ELF-int}) describes the motion of the vector
multiplet. The linear in the fields term in equations corresponds to
the "triply massless" ECS theory. The coupling includes
self-interaction term (\ref{EL-int}), and extra contribution
involving scalar field. It is convenient to think that
(\ref{ELF-int}) follows from (\ref{EL-int}) after the formal
redefinition of the model parameters
\begin{equation}\label{a-res}
    \phantom{\frac12}\alpha_1=0\,,\qquad \alpha_2=
    \gamma{}^2\phi{}^2\,,\qquad \alpha_3=
    \widetilde{\gamma}{}^2\phi{}^2-1\,.\phantom{\frac12}
\end{equation}
The characteristic polynomial (\ref{cp}) for the model reads
\begin{equation}\label{}
    \phantom{\frac12}M(\alpha;z)=\gamma^2\phi^2z^2+(\widetilde{\gamma}^2\phi^2-1)z^3\,.\phantom{\frac12}
\end{equation}
If the scalar field is set to nonzero constant from outset, we
obtain theory (\ref{EL-int}) with the second-order resonance for
zero root (case (2a) of classification on pages 4-5). As it has been
explained above, the corresponding model describes a massless spin-1
field, and a massive spin 1 vector subjected to the self-duality
condition. It is stable at the classical and quantum levels. The
second equation in the system (\ref{ELF-int}) describes the motion
of scalar field. This equation includes $\phi^3$-coupling, which
ensures the existence of non-zero stationary solution for
$\mathcal{A}_\mu=0$. This means that $\phi$ serves as the Higgs
field in the model (\ref{ELF-int}).

The interaction, being defined by equations (\ref{ELF-int}), is
consistent. The equations are preserved by the standard Yang-Mills
gauge symmetry (\ref{gt-int}). The scalar field is preserved by the
gauge transformation. The gauge identity has a slightly different
form,
\begin{equation}\label{gid-F-int}
    \mathbb{D}_\mu\mathbb{T}^\mu=0\,,\qquad
    \mathbb{D}_\mu=D_\mu+\frac{\widetilde{\gamma}{}^2\phi{}^2-1}{\beta_1{}^2}
    [\beta_1\mathcal{F}{}_\mu+\beta_2\mathcal{G}{}_\mu,\cdot]\,.
\end{equation}
We note that the gauge generator involves the scalar field
explicitly. The model has the same number of physical degrees of
freedom as the free theory because the orders of equations of motion
(\ref{ELF-int}), gauge identities (\ref{gid-F-int}), and gauge
symmetries (\ref{gt-int}) are preserved by coupling. Equations
(\ref{ELF-int}) do not follow from the least action principle from
the least action principle for a functional with higher derivatives
unless $\beta_2=0$. In the case of Lagrangian coupling, the action
principle reads
\begin{equation}\label{}
    S[\phi(x),\mathcal{A}(x)]=\frac{1}{2}\int\Big\{\mathcal{F}_\mu(\mathcal{G}^\mu+\gamma{}^2\phi^2\mathcal{F}^\mu)+\partial_\mu\phi\partial^\mu\phi+m^2\phi^2-\frac{1}{2}\phi^4\Big\}d^3x\,.
\end{equation}
The dynamics of the Lagragian theory is unstable. In the case of
non-Lagragian couplings, the dynamics of the model can be stable. In
the remaining part of the section, we address an issue of
construction of the constrained Hamiltonian formalism with a bounded
Hamiltonian for the higher derivative equations (\ref{ELF-int}).
Existence of such a formalism implies classical and quantum
stability of the model. To avoid "tautology" we outline to the most
crucial steps of the construction, while the details are provided
above.

We introduce a special notation for the linear combination of
generalized strengths $\mathcal{F}{}_\mu$, $\mathcal{G}{}_\mu$
entering the free part equations (\ref{ELF-int}),
\begin{equation}\label{W-def}
    \mathcal{W}{}_\mu=(\widetilde{\gamma}{}^2\phi{}^2-1)\mathcal{G}{}{}_\mu+\gamma^2\phi{}^2\mathcal{F}{}_\mu\,.
\end{equation}
We chose the space components of the vectors $\mathcal{F}{}_i$,
$\mathcal{W}{}_i$, $i=1,2$ (\ref{FiGi}) as variables absorbing first
and second derivatives of the original dynamical field
$\mathcal{A}{}_\mu$. In the sector of scalar field, we introduce the
canonical momentum $\pi=\dot{\phi}$. We consider $\mathcal{W}{}_0$
as a special notation for the combination (\ref{W-def}) of the
derivatives of phase-space variables
$\mathcal{A}{}_\mu,\mathcal{F}{}_i$. In terms of the variables
$\phi$, $\pi$ $\mathcal{A}{}_\mu$, $\mathcal{F}{}_i$,
$\mathcal{G}{}_i$, the first-order equations eventually read
\begin{equation}\label{dotA-res}
    \phantom{\frac12}\dot{\mathcal{A}}{}_i=\partial_i\mathcal{A}_0-
    [\mathcal{A}{}_0,\mathcal{A}{}_i]-\epsilon_{ij}\mathcal{F}_j\,;\phantom{\frac12}
\end{equation}
\begin{equation}\label{dotF-res}
    \phantom{\sum_1^2}\dot{\mathcal{F}}{}_i=\partial_i\mathcal{F}_0+[\mathcal{A}{}_i,\mathcal{F}{}_0]-
    [\mathcal{A}{}_0,\mathcal{F}{}_i]-
    \epsilon_{ij}\frac{\mathcal{W}{}_j-\gamma{}^2\phi{}^2\mathcal{F}{}_j}{\widetilde{\gamma}^2\phi^2-1}\,;\phantom{\sum_1^2}
\end{equation}
\begin{equation}\label{dotG-res}\begin{array}{c}\displaystyle
   \dot{\mathcal{W}}{}_i=\partial_i\mathcal{W}{}_0+[\mathcal{A}{}_i,\mathcal{W}{}_0]-[\mathcal{A}{}_0,\mathcal{W}{}_i]+
   \epsilon_{ij}[\beta_2\mathcal{W}{}_0-\beta_1\mathcal{F}{}_0,
   \beta_2\mathcal{W}{}_j-\beta_1\mathcal{F}{}_j]\,;\phantom{\sum_1^2}
\end{array}
\end{equation}
\begin{equation}\label{dotpi-res}\begin{array}{c}\displaystyle
    \phantom{\sum_1^2}\dot{\pi}=\Big\{\partial_i\partial_i-
    m^2+\widetilde{\gamma}{}^2\frac{(\beta_2\mathcal{W}{}_\mu-\beta_1\mathcal{F}_\mu)
    (\beta_2\mathcal{W}{}^\mu-\beta_1\mathcal{F}^\mu)}{(\widetilde{\gamma}{}^2\phi{}^2-1)^2}+\phi^2\Big\}\phi\,;\phantom{\sum_1^2}
\end{array}\end{equation}
\begin{equation}\label{dotphi-res}
    \phantom{\sum_1^2}\dot{\phi}=\pi
    \,.\phantom{\sum_1^2}
\end{equation}
The evolutionary equations (\ref{dotA-res}), (\ref{dotF-res}),
(\ref{dotG-res}), (\ref{dotpi-res}), (\ref{dotphi-res}) are
supplemented by the constraint
\begin{equation}\label{T-const-res}\begin{array}{c}\displaystyle
    \Theta=\epsilon_{ij}\Big(\partial_i\mathcal{W}{}_j+
    [\mathcal{A}{}_i,\mathcal{W}{}_j]+\frac{1}{2\beta_2{}^2}
    [\beta_2\mathcal{W}{}_i-\beta_1\mathcal{F}{}_i,
    \beta_2\mathcal{W}{}_j-
    \beta_1\mathcal{F}{}_j]\Big)=0\,.
\end{array}\end{equation} The constraint conserves on-shell if
the identity (\ref{gid-F-int}) is taken into account. The higher
derivative system (\ref{ELF-int}) follows from equations
(\ref{dotA-res}), (\ref{dotF-res}), (\ref{dotG-res}),
(\ref{T-const-res}) if all the extra variables are expressed in
terms of derivatives of original dynamical fields
$\phi,\mathcal{A}$.

The model (\ref{ELF-int}) admits a second-rank symmetric conserved
tensor,

\begin{equation}\label{}\begin{array}{c}\displaystyle
    \boldsymbol{\Theta{}^{\mu\nu}}(\beta;\alpha)=\text{tr}\Big\{\frac{\widetilde{\gamma}^2\phi^2-1}{\beta_2}
    \Big((\beta_1\mathcal{F}^{\mu}+\beta_2\mathcal{G}^\mu)(\beta_1\mathcal{F}^{\nu}+\beta_2\mathcal{G}^\nu)-
    \frac{}{2}g^{\mu\nu}(\mathcal{F}{}_{\rho}+\mathcal{G}{}_\rho)(\mathcal{F}{}^{\rho}+\mathcal{G}{}^\rho)\Big)+
    \\[5mm]\displaystyle+\frac{\beta_2{}^2}{\beta_1}\Big(\mathcal{F}{}^\mu\mathcal{F}{}^\nu-\frac{1}{2}g^{\mu\nu}\mathcal{F}{}_\rho
    \mathcal{F}{}^\rho\Big)\Big\}+\partial^\mu\phi\partial^\nu\phi+g^{\mu\nu}(-\frac{1}{2}\partial_\rho\phi\partial^\rho\phi-
    \frac{1}{2}m^2\phi^2+\frac{1}{4}\phi^4)\,;
\end{array}\end{equation}
\begin{equation}\label{}
    \phantom{\sum_1^2}\partial_\nu\boldsymbol{\Theta{}^{\mu\nu}}(\beta;\alpha)=-\text{tr}(\epsilon^{\mu\nu\rho}
    (\beta_1\mathcal{F}{}_\nu+\beta_2\mathcal{G}{}_\nu)\mathbb{T}{}_\rho)+\partial^\mu\phi\cdot\mathbb{T}\,.\phantom{\sum_1^2}
\end{equation}
The Hamiltonian is given by a sum of the $00$-component of conserved
tensor, being expressed in terms of variables $\phi$, $\pi$,
$\mathcal{A}{}_\mu$, $\mathcal{F}{}_i$, $\mathcal{W}{}_i$
(\ref{FiGi}), (\ref{W-def}), and a constraint term,
\begin{equation}\label{}\begin{array}{c}\displaystyle
    \mathcal{H}=\text{tr}\Big\{\frac{(\beta_2\mathcal{W}{}^0-\beta_1\mathcal{F}{}^0)(\beta_2\mathcal{W}{}^0-\beta_1\mathcal{F}{}^0)+
    (\beta_2\mathcal{W}{}^i-\beta_1\mathcal{F}{}^i)(\beta_2\mathcal{W}{}^i-\beta_1\mathcal{F}{}^i)}{2(\widetilde{\gamma}^2\phi^2-1)\beta_2}
    +\\[5mm]
    \displaystyle+\frac{\beta_1{}^2}{2\beta_2}(\mathcal{F}{}^0\mathcal{F}{}^0+\mathcal{F}{}^i
    \mathcal{F}{}^i)-(\mathcal{A}{}_0-
    \beta_1\mathcal{F}_0+\beta_2\mathcal{W}_{0})\Theta\Big\}+\\[5mm]\displaystyle\phantom{\frac12}
    +\frac{1}{2}(\pi\pi+
    \partial^i\phi\partial^i\phi)-
    \frac{1}{2}m^2\phi^2+\frac{1}{4}\phi^4
    \,.\phantom{\frac12}
\end{array}\end{equation}
Here, the quantities $\mathcal{F}{}_0,\mathcal{W}{}_0$ denote
abbreviations (\ref{F0G0}), (\ref{W-def}), and the numbers
$m,\gamma$ are model parameters. The Hamiltonian is on-shell bounded
if
\begin{equation}\label{}
    \widetilde{\gamma}{}^2\phi^2-1>0\,,\qquad \beta_2>0\,.
\end{equation}
These conditions involve the scalar field $\phi$. Once the initial
value of $\phi$ is a Cauchy data for equations (\ref{ELF-int}), the
Hamiltonian cannot be globally bounded. However, the Hamiltonian is
given by a positive definite quadratic form in the variables
$\mathcal{F}$, $\mathcal{G}$ in the range $|\widetilde\gamma\phi|>1$
of values of scalar field. This corresponds to the case of stability
island.

The Poisson bracket between the fields $\phi$, $\pi$,
$\mathcal{A}{}_\mu$, $\mathcal{F}{}_i$, $\mathcal{G}{}_i$ is
determined by the following system of equations:
\begin{equation}\label{PB-dotA-res}
    \phantom{\frac12}\Big\{\mathcal{A}{}_i,\int \mathcal H(\boldsymbol{y})
    d\boldsymbol{y}\Big\}\approx
    \partial_i\mathcal{A}_0-[\mathcal{A}{}_0,\mathcal{A}{}_i]-\epsilon_{ij}\mathcal{F}_j\,;\phantom{\frac12}
\end{equation}
\begin{equation}\label{PB-dotF-res}\begin{array}{c}\displaystyle
    \phantom{\frac12}\Big\{\mathcal{F}{}_i,\int \mathcal H(\boldsymbol{y}) d\boldsymbol{y}\Big\}\approx
    \partial_i\mathcal{F}_0+
    [\mathcal{A}{}_i,\mathcal{F}{}_0]-[\mathcal{A}{}_0,\mathcal{F}{}_i]-
    \epsilon_{ij}\frac{\mathcal{W}{}_j-\gamma{}^2\phi{}^2\mathcal{F}{}_j}{\widetilde{\gamma}{}^2\phi{}^2-1}\,;\phantom{\sum_1^2}
\end{array}\end{equation}
\begin{equation}\label{PB-dotG-res}\begin{array}{c}\displaystyle
    \Big\{G{}_i,\int \mathcal H(\boldsymbol{y}) d\boldsymbol{y}\Big\}
    \approx\partial_i\mathcal{W}{}_0+[\mathcal{A}{}_i,\mathcal{W}{}_0]-[\mathcal{A}{}_0,\mathcal{G}{}_i]
    +\frac{\epsilon_{ij}}{\beta_1{}^2}
    [\beta_2\mathcal{W}{}_0-\beta_1\mathcal{F}{}_0,\beta_2\mathcal{W}{}_j-\beta_1\mathcal{F}{}_j]\,.
\end{array}\end{equation}
\begin{equation}\label{PB-dotpi-res}\begin{array}{c}\displaystyle
    \phantom{\sum_1^2}\Big\{\pi,\int \mathcal H(\boldsymbol{y})
    d\boldsymbol{y}\Big\}\approx\partial_i\partial_i\phi+
    \Big(m^2-\widetilde{\gamma}^2
    \frac{(\beta_2\mathcal{W}{}_\mu-\beta_1\mathcal{F}{}_\mu)(\beta_2\mathcal{W}{}^\mu-\beta_1\mathcal{F}{}^\mu)}
    {(\widetilde{\gamma}^2\phi^2-1){}^2}
    -\phi^2\Big)\phi\,;\phantom{\sum_1^2}
\end{array}\end{equation}
\begin{equation}\label{PB-dotphi-res}
    \Big\{\phi,\int \mathcal
    H(\boldsymbol{y})d\boldsymbol{y}\Big\}\approx\pi.
\end{equation}
All the equalities are considered modulo terms that vanish modulo
constraint (\ref{T-const-res}). The solution for the equations
(\ref{PB-dotA-res}), (\ref{PB-dotF-res}), (\ref{PB-dotG-res}),
(\ref{PB-dotpi-res}), (\ref{PB-dotphi-res}) reads
\begin{equation}\label{PB-GG-FG-AA-res}
    \phantom{\sum_a^b}\phantom{\sum_a^b}\{\mathcal{W}{}^a{}_i(\boldsymbol{x}),
    \mathcal{W}{}^b{}_j(\boldsymbol{y})\}=\{\mathcal{F}{}^a{}_i(\boldsymbol{x}),
    \mathcal{W}{}^b{}_j(\boldsymbol{y})\}=\{\mathcal{A}{}^a{}_i(\boldsymbol{x}),
    \mathcal{F}{}^b{}_j(\boldsymbol{y})\}=0\,;\phantom{\sum_a^b}
\end{equation}
\begin{equation}\label{PB-FF-AG-res}
    \phantom{\sum_a^b}\phantom{\sum_a^b}\{\mathcal{A}{}^a{}_i(\boldsymbol{x}),
    \mathcal{W}{}^b{}_j(\boldsymbol{y})\}=-\{\mathcal{F}{}^a{}_i(\boldsymbol{x}),
    \mathcal{F}{}^b{}_j(\boldsymbol{y})\}=\frac{1}{\beta_1}\epsilon_{ij}\delta^{ab}
    \delta^{(2)}(\boldsymbol{x}-\boldsymbol{y})\,;\phantom{\sum_a^b}
\end{equation}
\begin{equation}\label{PB-AF-res}
    \phantom{\sum_a^b}\phantom{\sum_a^b}\{\mathcal{A}{}^a{}_i(\boldsymbol{x}),
    \mathcal{F}{}^b{}_j(\boldsymbol{y})\}=\frac{\beta_2}{\beta_1{}^2}\epsilon_{ij}\delta^{ab}
    \delta^{(2)}(\boldsymbol{x}-\boldsymbol{y})\,;\phantom{\sum_a^b}
\end{equation}
\begin{equation}\label{PB-phipi-res}
    \phantom{\sum_a^b}\{\phi(\boldsymbol{x}),
    \pi(\boldsymbol{y})\}=\delta{}^{(2)}(\boldsymbol{x}-\boldsymbol{y})\,.\phantom{\sum_a^b}
\end{equation}
All the Poisson brackets between $\phi$, $\pi$ and
$\mathcal{A}{}_i$, $\mathcal{F}{}_i$, $\mathcal{G}_i$ vanish. The
relations (\ref{PB-GG-FG-AA-res}), (\ref{PB-FF-AG-res}),
(\ref{PB-AF-res}) can be obtained from (\ref{PB-GG}), (\ref{PB-FG}),
(\ref{PB-FF}), (\ref{PB-AF}) after substitution (\ref{a-res}). We
note that the Poisson bracket between the fields $\mathcal{A}{}_i$,
$\mathcal{F}{}_i$, $\mathcal{W}{}_i$ is constant, even though the
theory involves extra fields $\pi,\phi$.

Now, we can demonstrate the phenomenon of dynamics stabilization by
means of "Higgs-like" mechanism proposed in \cite{KLN20}. Equations
(\ref{ELF-int}) admit a nonzero stationary solution
\begin{equation}\label{vac-res}
    \phantom{\sum_a^b}\mathcal{A}(\boldsymbol{x})=\mathcal{F}(\boldsymbol{x})=\mathcal{G}(\boldsymbol{x})=0\,,\qquad\pi(\boldsymbol{x})=0\,,\qquad
    \phi(\boldsymbol{x})=\pm m\,.\phantom{\sum_a^b}
\end{equation}
Introducing notation
$\phi^\ast(\boldsymbol{x})=\phi(\boldsymbol{x})\mp m$ and expanding
equations (\ref{ELF-int}) in the vicinity of this vacua, we obtain
the linearized equations for the fields
\begin{equation}\label{}\begin{array}{c}\displaystyle
    \phantom{\sum_a^b}\mathbb{T}{}_\mu=(m{}^2\widetilde{\gamma}{}^2-1)\mathcal{K}{}_\mu+m{}^2\gamma{}^2\mathcal{G}{}_\mu+\ldots=0\,,\phantom{\sum_a^b}\\[5mm]\displaystyle
    \phantom{\sum_a^b}\mathbb{T}=(\partial_\mu\partial^\mu+5m{}^2)\phi{}^\ast+\ldots=0\,.\phantom{\sum_a^b}
\end{array}\end{equation} The dots denote the terms that are at least
quadratic in the fields. As we see, the dynamics of the field
$\mathcal{A}$ is described by the higher derivative equation
(\ref{EL-free}) with $\alpha_3=m{}^2\widetilde{\gamma}{}^2-1$,
$\alpha_2=m{}^2\gamma{}^2$, $\alpha_1=0$. In this cases, the system
has a second order resonance for zero root, which does not affect
the stability of dynamics, at least at the free level. The on-shell
Hamiltonian reads
\begin{equation}\label{H-res-ast}\begin{array}{c}\displaystyle
    \mathcal{H}=\text{tr}\Big\{\frac{1}{2(m{}^2\widetilde{\gamma}{}^2-1)\beta_2}
    ((\beta_1\mathcal{W}^0-\beta_1\mathcal{F}^{0})(\beta_1\mathcal{W}^0-\beta_1\mathcal{F}^{0})+
    (\beta_2\mathcal{W}{}^i-\beta_1\mathcal{F}{}^i)(\beta_2\mathcal{W}{}^i-\beta_1\mathcal{F}{}^i))+\\[5mm]
    \displaystyle+\frac{\beta_1{}^2}{2\beta_2}(\mathcal{F}{}^0\mathcal{F}{}^0+\mathcal{F}{}^i
    \mathcal{F}{}^i)-(\mathcal{A}{}_0-\beta_1\mathcal{F}_0+\beta_2\mathcal{W}_{0})\Theta\Big\}+\\[5mm]\displaystyle
    +\frac{1}{2}(\pi\pi+\partial{}^i\phi{}^\ast\partial{}^i\phi{}^\ast+5m{}^2(\phi{}^\ast){}^2)+\ldots\phantom{\frac12}
\end{array}\end{equation}
The dots denote the terms that are at least cubic in the fields. The
quadratic part of the Hamiltonian is on-shell bounded if
$m\widetilde{\gamma}>1$. This means that the dynamics of small
fluctuations in the vicinity of vacua (\ref{vac-res}) is stable.
Once local stability is sufficient for construction of stable
quantum theory, the "Higgs-like" mechanism can stabilize the
dynamics of the non-linear theory (\ref{EL-int}) with the
third-order resonance.

\section{Conclusion}

In this article, we have studied the issue of stability of the most
general gauge ECS theory of third order for a vector multiplet. We
have shown that the free theory with $n$ dynamical fields admits a
$2n$-parameter series of symmetric conserved tensors of second rank.
The series includes a canonical energy-momentum. Even though the
canonical energy of the model is unbounded from below in all the
instances, the other conserved tensors have bounded $00$-component.
The stability conditions for dynamics are satisfied if the model has
no resonance (the characteristic equation has simple real roots, the
spectrum of masses is not degenerate). The only exception is the
possible multiplicity two zero root, which corresponds to the
Maxwell mode. We have confirmed the stability of dynamics by
construction of the non-canonical (non-Ostrogradski) Hamiltonian
formalism with the Hamiltonian, being given by the $00$-component of
a selected representative in the conserved tensor series. We have
observed that a bounded Hamiltonian is admitted by the model with no
resonance or order two resonance for the zero root.

At next step of our analysis, the self-interaction vertex with the
Yang-Mills gauge symmetry is proposed for the higher derivative
theory. The interaction preserves a selected conserved tensor of
free model, being determined by the values of coupling constants.
The Lagrangian interaction, which has been previously proposed in
\cite{Dai1}, leads to unstable dynamics. The couplings that preserve
the conserved tensors with bounded $00$-component are
non-Lagrangian. The non-linear dynamics admits a Hamiltonian form.
On-shell, the Hamiltonian is given by the $00$-component of the
conserved tensor. In case of Lagrangian interacting theory, the
Hamiltonian follows from the Ostrogradski procedure. In the case of
non-Lagrangian interactions, we have constructed an alternative
Hamiltonian formalism for a system of higher-derivative equations.
The alternative Hamiltonian is on-shell bounded if a conserved
tensor with the bounded $00$-component is preserved at the
interacting level. This observation demonstrates that the stability
of higher derivative dynamics can be consistent with non-abelian
gauge symmetry, even though the canonical energy of the model is
unbounded. To our knowledge the proposed theory (\ref{EL-int}) is a
first stable higher derivative model with non-abelian gauge
symmetry.

In the final part of the paper, we have considered the theory with
the third-order resonance, which can be called the "triply massless"
ECS model. This theory admits non-abelian interactions with
Yang-Mills gauge symmetry, but the $00$-component of the conserved
quantity is unbounded in all instances. To solve the problem, we
apply the "Higgs-like" mechanism of the paper \cite{KLN20}. An extra
scalar field with the $\phi^3$ (at the level of equations of motion)
self-coupling is introduced. The dynamics of the model admits the
Hamiltonian form, while the Hamiltonian has a local minimum for
nonzero value of the Higgs field. The dynamics of small fluctuations
in the vicinity of energy minimum is stable.

\vspace{0.5cm} \noindent {\bf Acknowledgments.} The authors thank
A.A. Sharapov for valuable discussions of this work and comments on
the manuscript. The part of the work concerning construction of
stable interactions in the ECS model (Sections 2,3) is supported by
the Ministry of Science and Higher Education of the Russian
Federation, Project No. 0721- 2020-0033. The construction of
Hamiltonian formalism (Sections 4,5) is worked out with the support
of the Russian Science Foundation grant 18-72-10123 in association
with the Lebedev Physical Institute of RAS. The work of SLL was
partially supported by the Foundation for the Advancement of
Theoretical Physics and Mathematics "BASIS".

\appendix

\section{Uniqueness of consistent interaction vertex}

In this Appendix, we demonstrate the fact of uniqueness of
interaction vertex (\ref{EL-int}) in the class of Poincare-covariant
couplings that are polynomial in the invariants $\mathcal{F}{}_\mu$,
$\mathcal{G}{}_\mu$ without higher derivatives. We apply the method
of inclusion of not necessarily Lagrangian consistent interactions
of article \cite{KLS-inv}.

Let us first explain the concept of interaction consistency in class
of non-Lagragian field theories. The dynamics of theory is
determined by a system of partial differential equations (equations
of motion) imposed onto the dynamical fields $\varphi^I(x)$,
\begin{equation}\label{T-eq}
    \phantom{\frac12}\mathbb{T}_a(\varphi^I(x),\partial\varphi^I(x),\partial^2\varphi^I(x),\ldots)=0\,.\phantom{\frac12}
\end{equation}
The equations of motion do not necessarily follow from the least
action principle for any functional $S[\varphi(x)]$, so $I$ and $a$
may run over different sets. For the free model, the left hand side
of (\ref{T-eq}) is supposed to be linear in the fields. The
interactions are associated with the deformations of system
(\ref{T-eq}) by non-linear terms. The equations of motion for the
theory with coupling are polynomial in the fields,
\begin{equation}\label{T-eq-int}
    \phantom{\frac12}\mathbb{T}{}_a=\mathbb{T}{}^{(0)}{}_a+\mathbb{T}{}^{(1)}{}_a
    +\mathbb{T}{}^{(2)}{}_a+\ldots=0\,.\phantom{\frac12}
\end{equation}
Here, $\mathbb{T}^{(0)}$, $\mathbb{T}^{(1)}$, and $\mathbb{T}^{(2)}$
are linear, quadratic, and cubic in the dynamical variables.
Throughout the section, the system (\ref{T-eq}) (or, equivalently,
(\ref{T-eq-int})) is supposed to be involutive. The concept of
involution implies that equations (\ref{T-eq}) have no differential
consequences of lower order (hidden integrability conditions). The
ECS equations (\ref{EL-int}) are involutive.

The defining relations for gauge symmetries and gauge identifies in
the system of partial derivative equations (\ref{T-eq}) read
\begin{equation}\label{gt-inv}
    \phantom{\frac12}\delta_\epsilon \varphi^I=R{}^I{}_\alpha\epsilon^\alpha,\qquad
    \delta_\epsilon \mathbb{T}_a|_{\mathbb{T}=0}=0\,;\phantom{\frac12}
\end{equation}
\begin{equation}\label{gid-inv}
    \phantom{\frac12}L{}^a{}_A\mathbb{T}_a\equiv0\,.\phantom{\frac12}
\end{equation}
Here $R{}^I{}_\alpha, L{}^a{}_A$ are certain differential operators.
For non-Lagrangian equations, the gauge symmetries and gauge
identities are not related to each other, so the multi-indices
$A,\alpha$ are different. In the perturbative setting, the gauge
symmetry and gauge identity generators are supposed to be polynomial
in fields
\begin{equation}\label{}
R{}^I{}_\alpha=R{}^{(0)}{}^I{}_\alpha+R{}^{(1)}{}^I{}_\alpha+
R{}^{(2)}{}^I{}_\alpha+\ldots\,,\qquad
L{}^a{}_A=L{}^{(0)}{}^a{}_A+L{}^{(1)}{}^a{}_A+
L{}^{(2)}{}^a{}_A+\ldots
\end{equation}
Here, $R^{(0)},L^{(0}$, $R^{(1)},L^{(1)}$, and $R^{(2)},L^{(2)}$ are
field-independent, linear, and quadratic in the dynamical variables.
The interaction is consistent if all the gauge symmetries and gauge
identities of free model are preserved by the deformation of
equations of motion.  Equation (8) of paper \cite{KLS-inv} provides
a simple formula for computation of physical degrees of freedom
number based on the orders of derivatives involved into equations of
models, gauge symmetries, and gauge identities. The free and
non-linear theory must have one and the same number of physical
degrees of freedom.

Relations (\ref{gt-inv}), (\ref{gid-inv}) imply the following
consistency conditions for $\mathbb{T}^{(k)}$, $R^{(k)}$, $L^{(k)}$,
$k=0,1,2,\ldots$:
\begin{equation}\label{R0}
    \phantom{\frac12}R^{(0)}{}^I{}_\alpha\partial_I \mathbb{T}{}^{(0)}{}_{a}=0\,;\phantom{\frac12}
\end{equation}
\begin{equation}\label{R1}
    \phantom{\frac12}R^{(0)}{}^I{}_\alpha\partial_I \mathbb{T}{}^{(1)}{}_{a}+R^{(1)}{}^I{}_\alpha\partial_I \mathbb{T}{}^{(0)}{}_{a}=0\,;\phantom{\frac12}
\end{equation}
\begin{equation}\label{R2}
    \phantom{\frac12}R^{(0)}{}^I{}_\alpha\partial_I \mathbb{T}{}^{(2)}{}_{a}+
    R^{(1)}{}^I{}_\alpha\partial_I \mathbb{T}{}^{(1)}{}_{a}+R^{(2)}{}^I{}_\alpha\partial_I \mathbb{T}{}^{(0)}{}_{a}=0\,
\phantom{\frac12}\end{equation} (the symbol $\partial_I$ denotes a
variational derivative with respect to the field $\varphi{}^I$);
\begin{equation}\label{L0}
    \phantom{\frac12}L^{(0)}{}^a{}_A \mathbb{T}{}^{(0)}{}_{a}=0\,;\phantom{\frac12}
\end{equation}
\begin{equation}\label{L1}
   \phantom{\frac12} L^{(0)}{}^a{}_A \mathbb{T}{}^{(1)}{}_{a}+L^{(1)}{}^a{}_A \mathbb{T}{}^{(0)}{}_{a}=0\,;\phantom{\frac12}
\end{equation}
\begin{equation}\label{L2}
    \phantom{\frac12}L^{(0)}{}^a{}_A \mathbb{T}{}^{(2)}{}_{a}+
    L^{(1)}{}^a{}_A \mathbb{T}{}^{(1)}{}_{a}+L^{(2)}{}^a{}_A
    \mathbb{T}{}^{(0)}{}_{a}=0\,.\phantom{\frac12}
\end{equation}
The equations (\ref{R0}), (\ref{L0}) determine the gauge symmetry
and gauge identity generators $R^{(0)}$, $L^{(0)}$ of the free
theory. These quantities are usually given from outset. Relations
(\ref{R1}), (\ref{L1}) determine the first-order corrections to the
equations of motion $T^{(1)}$, gauge symmetry generators $R^{(1)}$,
and gauge identity generators $L^{(1)}$. Relations (\ref{R1}),
(\ref{L1}) determine the second-order corrections to the equations
of motion $T^{(2)}$, gauge symmetry generators $R^{(2)}$, and gauge
identity generators $L^{(2)}$. The procedure of interaction
construction can be extended to the third and higher orders. Once
the most general covariant ansatz is applied for $T^{(k)}$,
$k=1,2,\ldots$ the procedure (\ref{R0}), (\ref{R1}), (\ref{R2}),
(\ref{L0}), (\ref{L1}), (\ref{L2}), $\ldots$ allows a complete
classification of consistent interactions in a given field theory.
An important subtlety of this procedure is that some lower-order
couplings can be inconsistent at the higher orders of perturbation
theory. The first critical step is the extension of the first-order
(quadratic) interaction vertex at the second-order of perturbation
theory.

Relation (\ref{EL-free}) determines the left hand side of the free
ECS equations $\mathbb{T}{}^{(0)}$,
\begin{equation}\label{T0}
    \phantom{\frac12}\mathbb{T}^{(0)}{}^a{}_\mu=\alpha_1F{}^{a}{}_\mu+\alpha_2G{}^{a}{}_\mu+\alpha_3K{}^{a}{}_\mu=0\,.\phantom{\frac12}
\end{equation}
The gauge symmetries and gauge identities are defined by the
gradient and divergence operators,
\begin{equation}\label{}
    \phantom{\frac12}R{}^{(0)}{}_\mu=\partial{}_\mu\,,\qquad
    L^{(0)}{}{}^\mu=\partial{}^\mu\,.\phantom{\frac12}
\end{equation}
The free gauge identity (\ref{R0}) and free gauge transformation
read
\begin{equation}\label{}
    \phantom{\frac12}\partial{}_\mu \mathbb{T}{}^{(0)}{}^\mu\equiv0\,,\qquad \delta_\epsilon
    A{}^a{}_\mu=\partial_\mu\zeta{}^a\,,\phantom{\frac12}
\end{equation}
where $\zeta$'s are gauge transformation parameters. We consider the
Poincare-covariant interactions such that are expressed in terms of
gauge covariants $\mathcal{F}{}_\mu$, $\mathcal{G}{}_\mu$,
$\mathcal{K}{}_\mu$ (\ref{FG}) with no higher-derivative terms being
included into coupling. In this case, the equations of motion are
automatically preserved by the Yang-Mills gauge symmetry (equations
(\ref{R0}), (\ref{R1}), (\ref{R2}) are satisfied). Consistent
interaction vertices of first and second orders are selected by the
conditions (\ref{L1}), (\ref{L2}). We elaborate on this problem
below.

We assume that the equations of motion are polynomial in gauge
covariants $\mathcal{F}{}_\mu$, $\mathcal{G}{}_\mu$,
$\mathcal{K}{}_\mu$. The linear term is given by the
covariantization of the free equations (\ref{EL-free}). The most
general covariant first-order interaction vertex without
higher-derivatives reads
\begin{equation}\label{T1-int}\begin{array}{c}\displaystyle
    \phantom{\frac12}\mathbb{T}{}^{(0)}{}_\mu+\mathbb{T}{}^{(1)}{}_\mu=\phantom{\frac12}
    \\[7mm]\displaystyle=
    \alpha_1\mathcal{F}{}_\mu+
    \alpha_2\mathcal{G}{}_\mu+\alpha_3\mathcal{K}{}_\mu+\epsilon_{\mu\nu\rho}
    \Big(\frac{1}{2}k{}_1[\mathcal{F}{}^\nu,\mathcal{F}{}^\rho]+
    k{}_2[\mathcal{F}{}^\nu,\mathcal{G}{}^\rho]+
    \frac{1}{2} k{}_3[\mathcal{G}{}^\nu,\mathcal{G}{}^\rho]\Big)\,,
\end{array}\end{equation}
where $k_l,l=1,2,3$ are constants. The covariant divergence of
equations of motion reads
\begin{equation}\label{}\begin{array}{c}\displaystyle
    D_\mu
    \mathbb{T}^\mu=-\frac{1}{\alpha_3}[k_2\mathcal{F}{}_\mu+k_3\mathcal{G}{}_\mu,\mathbb{T}{}^\mu]+\\[5mm]\displaystyle
    +\frac{1}{\alpha_3}(\alpha{}_3{}^2-\alpha_3k_1+\alpha_2{}k_2-\alpha_1{}k_3)
    [\mathcal{F}{}_\mu,\mathcal{G}^{\mu}]+(k_2^2-k_3k_1)\varepsilon_{\mu\nu\rho}[\mathcal{F}{}^\mu,[\mathcal{F}{}^\nu,\mathcal{G}{}^\rho]]\,.
\end{array}\end{equation}
The gauge identity is satisfied in the first-order approximation
(\ref{L1}) if the coefficients $k_l,l=1,2,3$ satisfy relation
\begin{equation}\label{}
    \phantom{\frac12}\alpha{}_3{}^2-\alpha_3k_1+\alpha_2{}k_2-\alpha_1{}k_3=0\,.\phantom{\frac12}
\end{equation}
The general solution to this equations reads
\begin{equation}\label{k-int}
    k_1=-\frac{\beta{}_1{}^2\alpha_3{}^2}{C(\beta;\alpha)}+\alpha{}_1\beta{}_3\,,\qquad
    k_2=-\frac{\beta{}_2{}\beta{}_1\alpha_3{}^2}{C(\beta;\alpha)}\,\qquad
    k_3=-\frac{\beta{}_2{}^2\alpha_3{}^2}{C(\beta;\alpha)}+\alpha{}_3\beta{}_1\,,
\end{equation}
where $\beta_1$, $\beta_2$, $\beta_3$ are coupling parameters. The
parameters $\beta_1,\beta_2$ determine the coupling vertex
(\ref{EL-int}) (two constants determine a single coupling because
the ratio $\beta_1/\beta_2$ is relevant). The constant $\beta_3$ is
responsible for another interaction vertex, which is consistent at
the first order of perturbation theory. The interaction vertex
(\ref{EL-int}) is self-consistent with no higher-order corrections
required for the equations of motion. The other coupling needs cubic
in the fields corrections to the equations of motion. To prove the
uniqueness of interaction (\ref{EL-int}) we should demonstrate that
the ansatz (\ref{T1-int}), (\ref{k-int}) is inconsistent at the
second order of perturbation theory for $\beta_3\neq0$.

The most general second-order covariant interaction vertex reads
\begin{equation}\label{}\begin{array}{c}\displaystyle
    \mathbb{T}{}^{(2)}{}_\mu=l_1[\mathcal{F}{}_\nu,[\mathcal{F}{}_\mu,\mathcal{F}{}^\nu]]+
    l_2[\mathcal{F}{}_\nu,[\mathcal{G}{}_\mu,\mathcal{F}{}^\nu]]+
    l_3[\mathcal{F}{}_\nu,[\mathcal{F}{}_\mu,\mathcal{G}{}^\nu]]+\\[7mm]\displaystyle+
    l_4[\mathcal{G}{}_\nu,[\mathcal{F}{}_\mu,\mathcal{G}{}^\nu]]
    +l_5[\mathcal{G}{}_\nu,[\mathcal{G}{}_\mu,\mathcal{F}{}^\nu]]+
    l_6[\mathcal{G}{}_\nu,[\mathcal{G}{}_\mu,\mathcal{G}{}^\nu]]\,,
\end{array}\end{equation}
where $l_p,p=\overline{1,6}$ are constants. The covariant divergence
of equations of motion reads (only cubic terms are written out)
\begin{equation}\label{}\tag{A20}\begin{array}{c}\displaystyle
    D_\mu\mathbb{T}{}^\mu=\frac{1}{\alpha_3}\epsilon_{\mu\nu\rho}[\mathcal{S}^{\mu\nu},\mathbb{T}^\rho]+
    \frac{1}{2}C_1(k;l)\varepsilon_{\mu\nu\rho}[\mathcal{F}{}^\mu,[\mathcal{F}{}^\nu,\mathcal{G}{}^\rho]]+
    \frac{1}{2}C_2(k;l)\varepsilon_{\mu\nu\rho}[\mathcal{G}{}^\mu,[\mathcal{G}{}^\nu,\mathcal{F}{}^\rho]]+
    \\[5mm]\displaystyle+
    \frac{l_1}{2}
    [\mathcal{F}{}_\mu,[\mathcal{F}{}_\nu, D^\mu\mathcal{F}{}^\nu+D^\nu\mathcal{F}{}^\mu]]+
    \frac{l_3-l_2}{2}[\mathcal{G}{}_\mu,[\mathcal{F}{}_\nu,
    D^\mu\mathcal{F}{}^\nu+D^\nu\mathcal{F}{}^\mu]]+
    \frac{2l_2-l_3}{2}[\mathcal{F}{}_\mu,[\mathcal{G}{}_\nu,
    D^\mu\mathcal{F}{}^\nu+D^\nu\mathcal{F}{}^\mu]]+
    \\[5mm]\displaystyle +\frac{l_5}{2}[\mathcal{G}{}_\mu,[\mathcal{G}{}_\nu,
    D^\mu\mathcal{F}{}^\nu+D^\nu\mathcal{F}{}^\mu]]+\frac{l_3}{2}[\mathcal{F}{}_\mu,[\mathcal{F}{}_\nu,
    D^\mu\mathcal{G}{}^\nu+D^\nu\mathcal{G}{}^\mu]]+\frac{l_5-l_4}{2}[\mathcal{F}{}_\mu,[\mathcal{G}{}_\nu,
    D^\mu\mathcal{G}{}^\nu+\\[5mm]\displaystyle
    +D^\nu\mathcal{G}{}^\mu]]+
    \frac{2l_4-l_5}{2}[\mathcal{G}{}_\mu,[\mathcal{F}{}_\nu,
    D^\mu\mathcal{G}{}^\nu+D^\nu\mathcal{G}{}^\mu]]+\frac{l_6}{2}[\mathcal{G}{}_\mu,[\mathcal{G}{}_\nu,
    D^\mu\mathcal{G}{}^\nu+D^\nu\mathcal{G}{}^\mu]]\,.
\end{array}\end{equation}
Here, the notation is used,

\begin{equation}\label{}\tag{A21}
    C_1(k;l)=k_2{}^2-k_1k_3-3l_1-\frac{\alpha_1}{\alpha_3}l_4-
    \frac{\alpha_1}{\alpha_3}l_5-\frac{\alpha_1}{\alpha_3}l_6\,,\qquad
    C_2(k;l)=3l_2+l_3-\frac{\alpha_2}{\alpha_3}l_5-\frac{\alpha_1}{\alpha_3}l_6\,;
\end{equation}
\begin{equation}\label{}\tag{A22}\begin{array}{c}\displaystyle
    \mathcal{S}^{\mu\nu}=l_3[\mathcal{F}^\mu,[\mathcal{F}{}^\nu,\cdot]]+
    l_4[\mathcal{G}^\mu,[\mathcal{F}{}^\nu,\cdot]]+
    l_6[\mathcal{G}^\mu,[\mathcal{G}{}^\nu,\cdot]]
    -l_4[[\mathcal{F}^\mu,\mathcal{F}{}^\nu],\cdot]-
\\[7mm]\displaystyle
    -l_5[[\mathcal{F}^\mu,\mathcal{G}{}^\nu],\cdot]-
    l_6[[\mathcal{G}^\mu,\mathcal{G}{}^\nu],\cdot]\,.
\end{array}\end{equation}
The interaction is consistent at the second order of perturbation
theory if the right hand side of this expression vanish modulo free
equations (\ref{EL-free}). The critical observation is that the
expressions of the form
$[\mathcal{X}_\mu\,,[\mathcal{Y}{}_\nu\,,D^\mu\mathcal{Z}^\nu+D^\nu\mathcal{Z}^\mu]]$,
where
$\mathcal{Y}{}_\mu,\mathcal{Z}{}_\mu,\mathcal{Y}{}_\mu=\mathcal{F}{}_\mu$
or $\mathcal{G}{}_\mu$ represent on-shell independent combinations
of fields and their derivatives. Once they have to vanish, we
conclude
\begin{equation}\label{}\tag{A23}
    k_2{}^2-k_3k_1=0\,,\qquad l_p=0\,,\qquad p=\overline{1,6}\,.
\end{equation}
The general solution to these equations has the form (\ref{k-int})
with $\beta_3=0$. With account of this fact, the interaction
(\ref{EL-int}) is unique in the class of covariant couplings without
higher derivatives.

\end{document}